\documentclass[conference]{IEEEtran}
\IEEEoverridecommandlockouts
\usepackage{cite}
\usepackage{amsmath,amssymb,amsfonts}
\usepackage{algorithmic}
\usepackage{graphicx}
\usepackage{textcomp}
\usepackage{xcolor}
\usepackage{epsfig}
\usepackage{epstopdf}
\usepackage{balance}
\usepackage{placeins}
\usepackage[printonlyused]{acronym}
\usepackage{siunitx}
\usepackage[linesnumbered,ruled,vlined]{algorithm2e}
\usepackage{float}
\usepackage{silence}
\acrodef{fh}[FH]{frequency hopping}

\newcommand{\FGR}[1]{Fig.~\ref{#1}}

\def\BibTeX{{\rm B\kern-.05em{\sc i\kern-.025em b}\kern-.08em
    T\kern-.1667em\lower.7ex\hbox{E}\kern-.125emX}}

\begin{document}

\title{Communication Modeling of Long-Distance Abscisic Acid Signaling in Plant Vascular Systems}

\author{\IEEEauthorblockN{Necati Kagan Erkek \IEEEauthorrefmark{1}, Hani Ballouz \IEEEauthorrefmark{2}, Radin Monshian Motlagh\IEEEauthorrefmark{1}}

\IEEEauthorblockA{\IEEEauthorrefmark{1} Telecommunications Engineering, Dept. of Electronic, Information and Bioengineering, Politecnico di Milano, Milan, IT} 
\IEEEauthorblockA{\IEEEauthorrefmark{2} Environmental Engineering, Dept. of Civil and Environmental Engineering, Politecnico di Milano, Milan, IT} Emails: \texttt{\{necatikagan.erkek, hani.ballouz, radin.monshian\}@mail.polimi.it} 
}

\maketitle
\begin{abstract}
Abscisic acid (ABA) is a central plant hormone for coordinating responses to drought, salinity, cold stress, pathogen attack, wounding, and developmental aging. This paper reviews the biological stimuli that increase ABA biosynthesis, the main production sites and pathways, and the long-distance movement of ABA through plant vascular tissues. It then discusses experimental quantification approaches, including gas-liquid chromatography with electron-capture detection and high-performance liquid chromatography with ultraviolet detection. Finally, the paper presents a molecular-communication-inspired model of ABA transport in which root-side ABA release is represented as a transmitter, the xylem pathway as a bounded channel, and soybean tissue as a receiver. MATLAB Brownian-motion simulations are used to evaluate the effects of released molecule quantity and receiver radius on the detected ABA signal. The results show that higher release quantities produce smoother and stronger reception trends, while larger receivers increase molecule-capture probability.
\end{abstract}

\begin{IEEEkeywords}
Abscisic Acid ABA, ABA Biosynthesis, Cold stress, Osmotic stress, Xylem
\end{IEEEkeywords}

\section{Introduction}
In the intricate dance between plants and their ever-changing environments, a key player emerges – Abscisic Acid (ABA). A phytohormone renowned for its pivotal role in mediating responses to environmental stressors, particularly drought, ABA orchestrates a complex cascade of physiological events aimed at ensuring a plant's survival under adverse conditions. This project embarks on a journey to unravel the mysteries surrounding ABA, specifically focusing on the long-distance transport of this critical phytohormone between plant organs. The quest to understand ABA's dynamics within the plant system necessitates an exploration of various facets, encompassing stimuli that trigger its production, the specific sites where it is synthesized, the modes through which it traverses the plant's intricate network, experimental methodologies for quantifying its presence, and the development of models to simulate and comprehend its transport dynamics.

ABA production is intricately linked to the plant's perception of environmental stressors. Unraveling the stimuli that trigger heightened ABA production represents a fundamental step in understanding the plant's stress response mechanism. As the project delves into this aspect, it seeks to elucidate the diverse range of environmental cues – be it water scarcity, salinity, or other stressors – that act as signals prompting the plant to initiate ABA biosynthesis. The spatial regulation of ABA synthesis is a critical determinant in the orchestration of stress responses. Investigating the specific cellular locales where ABA is produced forms an integral part of this project. By pinpointing the production sites within plant organs, researchers aim to uncover the spatial intricacies that govern ABA synthesis, shedding light on the cellular machinery that responds to stress signals.

One of the key decisions to be made is the mode of transport whether ABA utilizes the xylem, the conduit for water transport, or the phloem, responsible for nutrient transport. Through meticulous investigation, this project seeks to discern the preferred pathway ABA follows as it travels from its site of synthesis to distal plant organs. A comprehensive understanding of ABA transport necessitates the development and application of experimental methodologies for precise quantification. Researchers engaged in this project aim to identify and employ cutting-edge techniques, leveraging experimental data to quantify ABA concentrations in different plant organs. By doing so, they aspire to provide a quantitative foundation for unraveling the intricacies of ABA movement within the plant \cite{b1,b2,b3,b4,b5,b6,b7,b8,b9}. To complement experimental findings, the project envisions creation of models that simulate ABA transport dynamics. By incorporating data on stimuli, production sites, and experimental quantification, these models strive to offer a holistic perspective on the intricate interplay of factors influencing ABA movement within the plant. Such modeling endeavors contribute to a deeper comprehension of regulatory mechanisms governing long-distance ABA transport. 

\begin{figure}[!ht]
    \centering
    \includegraphics[width=0.9\linewidth]{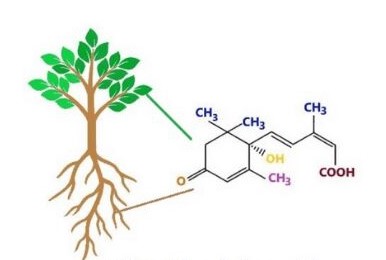}
    \caption{ABA in Plant}
    \label{fig1}
\end{figure}

\vspace{4pt}

In summary, this project embarks on a multidimensional exploration of Abscisic Acid, offering a comprehensive overview of the stimuli governing its production, the sites where it is synthesized, the vascular routes it employs for long-distance transport, experimental methodologies for quantification, and the modeling approaches that bring these facets together. Through these collective efforts, the aim is to contribute significantly to the evolving narrative of plant stress responses and the pivotal role played by ABA in orchestrating these intricate physiological maneuvers.

\section{What is Abscisic Acid?}
\subsection{Abscisic Acid (ABA) Overview}

Abscisic Acid (ABA) stands as a versatile and multifaceted signaling molecule, transcending its traditional role as a plant hormone. Extending its influence across diverse organisms, ABA has become a pivotal player in orchestrating various biological processes. Originally recognized for its crucial involvement in plant responses to environmental stresses, such as drought and salinity, ABA's reach has expanded to encompass an array of functions.

In plants, ABA acts as a regulatory force in stress adaptation, influencing critical processes like stomatal closure to conserve water during drought conditions. Moreover, it exerts control over seed development, dormancy, and root architecture, ensuring plants' resilience in changing environmental conditions. Beyond the botanical realm, recent research hints at ABA's involvement in non-plant organisms, suggesting roles in immune responses and stress adaptation in animals, including humans \cite{b11}.

As scientific exploration delves deeper into the molecular mechanisms of ABA signaling, new insights continually emerge, revealing its intricate roles in biological processes. The study of ABA stands as a dynamic field, where ongoing research not only enhances our comprehension of plant physiology but also broadens the perspective on the broader significance of ABA across various organisms and their adaptive strategies in response to environmental challenges \cite{b12}.

\begin{figure}[!ht]
    \centering
    \includegraphics[width=0.8\linewidth]{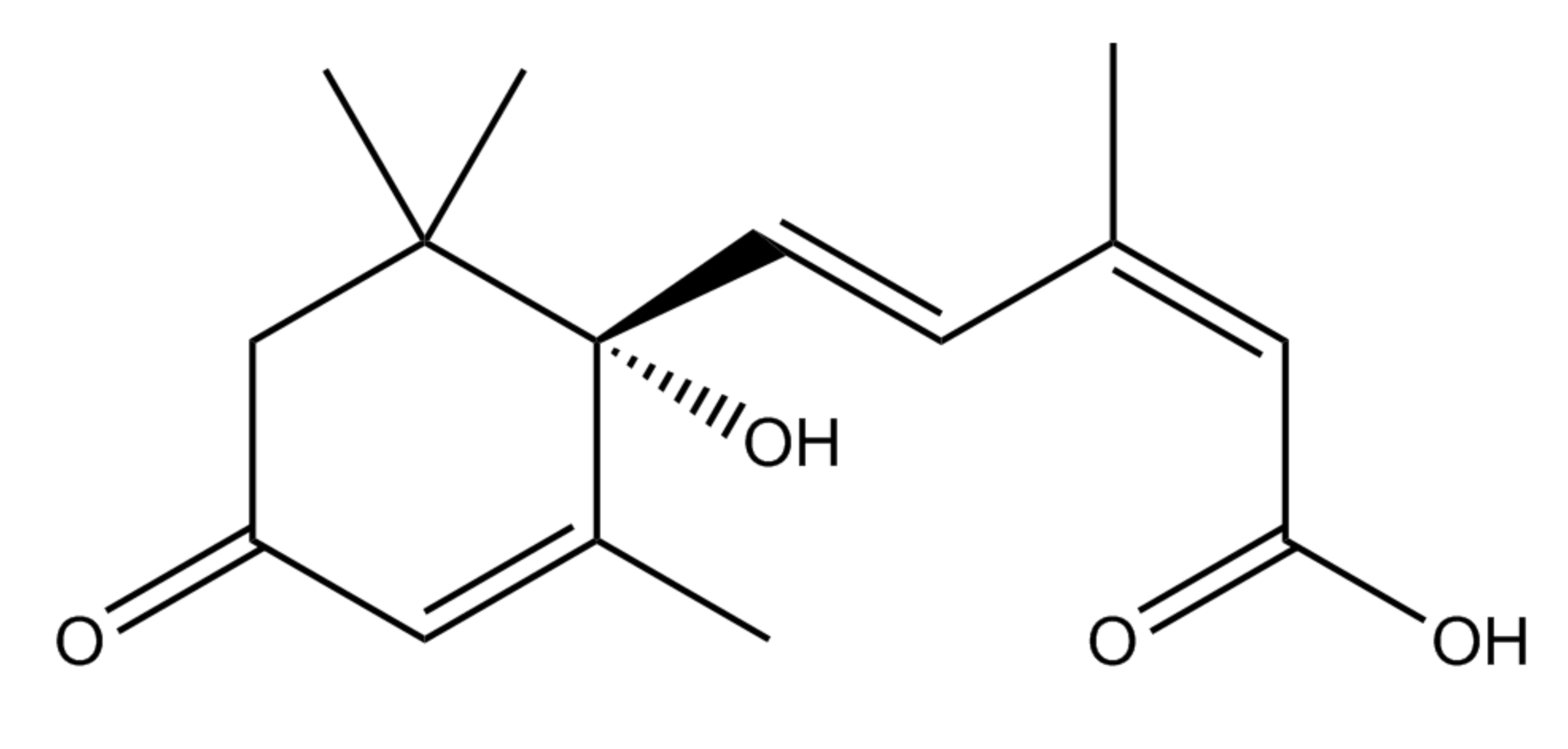}
    \caption{Abscisic Acid}
    \label{fig2}
\end{figure}

\begin{figure}[!ht]
    \centering
    \includegraphics[width=0.9\linewidth]{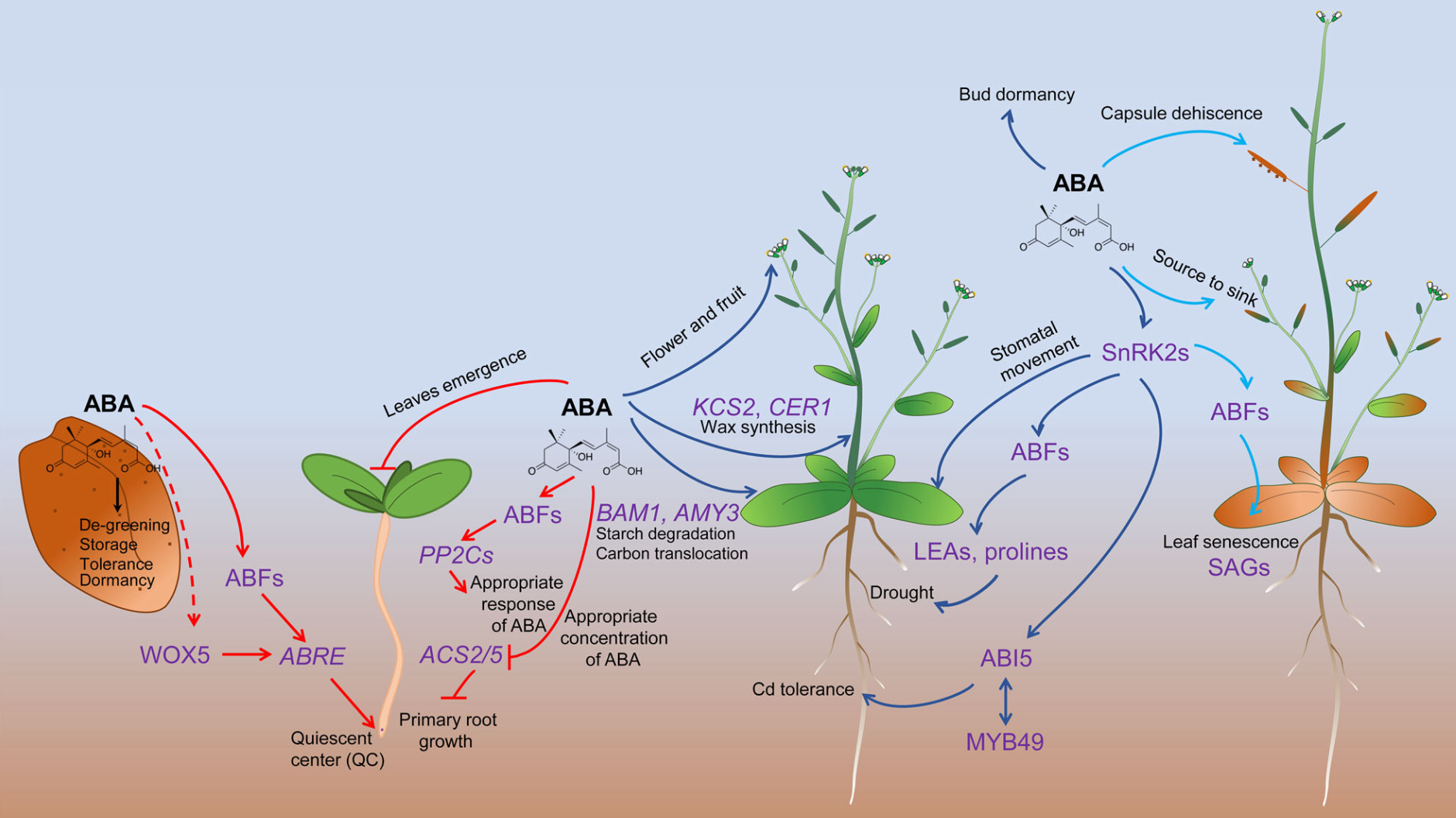}
    \caption{ABA Environment}
    \label{fig3}
\end{figure}

\subsection{Evolutionary Presence and Interactions}

Abscisic Acid (ABA) exhibits remarkable evolutionary conservation, being synthesized across diverse kingdoms of life. Its presence is not confined to plants but extends to cyanobacteria and even human cells, underscoring its ancient and widespread existence. The conservation of ABA throughout evolution highlights its fundamental roles and adaptive significance across different organisms. The fact that ABA, a plant hormone traditionally associated with stress response and growth regulation, is found in organisms as varied as cyanobacteria and human cells suggests its involvement in fundamental biological processes. This evolutionary persistence implies that ABA's signaling functions have been honed over millions of years, allowing organisms from various lineages to utilize its versatile properties for survival and adaptation. The study of ABA's presence and functions across diverse life forms contributes to our understanding of shared molecular mechanisms and adaptive strategies that have stood the test of evolutionary time.

\subsection{Medicinal Potential}

Abscisic Acid (ABA), primarily known for its pivotal role in plant physiology, is increasingly gaining attention for its potential medicinal applications in animals and humans. Recent research suggests that ABA may offer therapeutic benefits, particularly in the context of stomatal defense against biotic and drought stresses. The molecule's ability to regulate stress responses and immune mechanisms in plants has sparked interest in exploring its analogous functions in animals, hinting at potential applications in human medicine. ABA's capacity to modulate immune responses and its involvement in stress adaptation may have implications for developing novel therapeutic strategies. While the field is in its early stages, the promising attributes of ABA open avenues for future research and development, raising the prospect of clinical applications that harness the versatile signaling molecule for addressing various health challenges in both animals and humans.

\subsection{Kingdom-Specific Features}
Kingdom-specific features in the biosynthesis and signaling pathways of Abscisic Acid (ABA) highlight its remarkable adaptability across diverse biological contexts. ABA, a pivotal plant hormone, exhibits variations in its synthesis and signaling mechanisms, tailoring its functions to the specific needs of different kingdoms. The intricate interplay of ABA with distinct molecular components in various organisms underscores its versatility. Additionally, the identification of naturally occurring ABA in different stereoisomeric forms introduces an extra layer of complexity to its kingdom-specific functions. This structural diversity may contribute to the nuanced responses of ABA within different biological systems. Understanding these kingdom-specific features enhances our grasp of ABA's roles beyond plants and sheds light on its multifaceted functions across the evolutionary spectrum, emphasizing the dynamic nature of this signaling molecule in adapting to the unique challenges posed by different organisms.

\subsection{Functions}
Abscisic Acid (ABA) serves multifaceted functions crucial for plant survival and adaptation. Primarily, it induces seed dormancy and inhibits germination, especially post-maturation and in late embryogenesis. This strategic delay in germination ensures that seeds only sprout under optimal conditions, enhancing the plant's chances of successful growth. ABA's role in stomatal regulation is pivotal, triggering stomatal closure in response to water scarcity to curtail transpiration and conserve water. During abiotic stresses like drought, 
ABA acts as a signaling molecule, activating the expression of ABA-responsive genes. This molecular response orchestrates various adaptive mechanisms, aiding the plant in coping with adverse environmental conditions. The intricate interplay of ABA in seed development, stomatal regulation, and stress response showcases its versatility as a signaling molecule, highlighting its paramount importance in the plant's ability to thrive in diverse environmental scenarios.

\subsection{Transport}
Abscisic Acid (ABA) plays pivotal roles in plant physiology, with its functions extending to transport mechanisms critical for its distribution within the organism. ABA transport is a finely tuned process, employing both carrier protein-mediated active movement and pH-dependent passive movement. The intricate network of ABA transporters includes various types, such as ABA efflux carriers responsible for exporting ABA out of cells, and importers facilitating its entry. These transporters contribute to the dynamic regulation of ABA levels in different cellular compartments, allowing plants to respond effectively to environmental stimuli. The balance between active and passive transport mechanisms ensures the precise spatial and temporal control of ABA, influencing crucial processes like stomatal closure, seed dormancy, and stress responses. Understanding the intricacies of ABA transport sheds light on the sophisticated regulatory systems that plants employ to adapt and thrive in diverse environmental conditions.

\begin{figure}[!ht]
    \centering
    \includegraphics[width=0.75\linewidth]{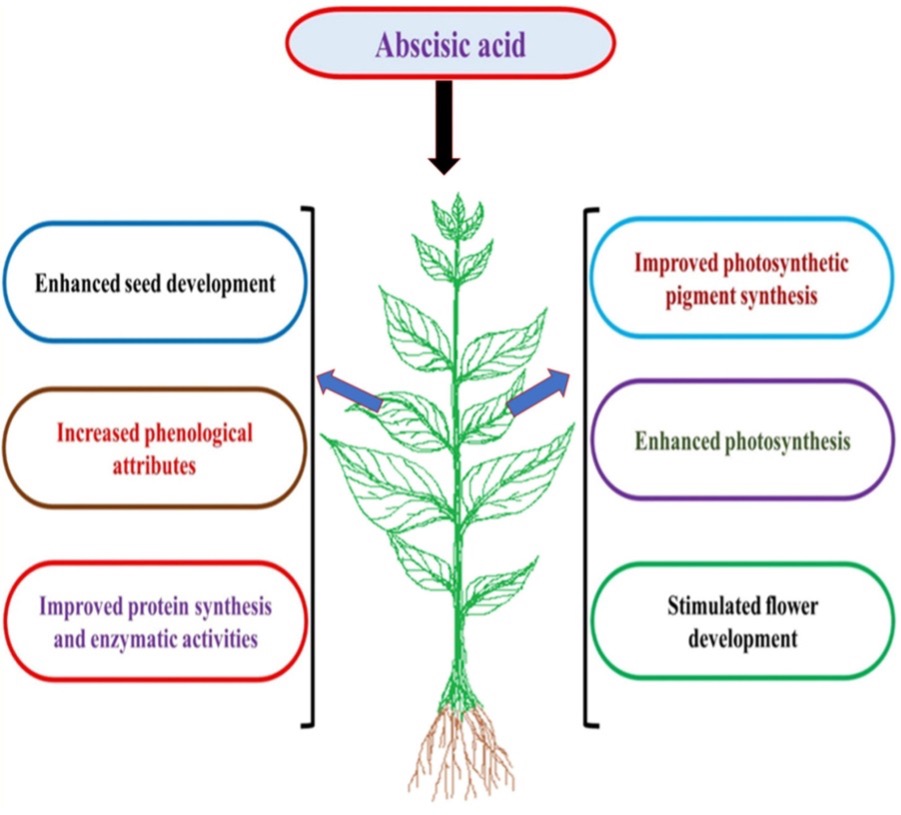}
    \caption{Abscisic Acid}
    \label{fig4}
\end{figure}

\subsection{Regulation of Gene Expression}
Abscisic Acid (ABA) exerts its diverse functions in plants through the intricate regulation of gene expression, particularly in response to environmental stresses. ABA plays a pivotal role in the transcriptional regulation of numerous genes, especially those involved in stress responses. The molecular mechanism involves the activation of SnRK2 protein kinases, which catalyze the phosphorylation of basic leucine zipper (bZIP) transcription factors. These phosphorylated bZIP factors, in turn, bind to ABA-responsive elements (ABRE) in the promoter regions of target genes. This binding initiates a cascade of events leading to the activation or repression of gene expression, orchestrating adaptive responses to various environmental challenges like drought or high salinity. The ABA-mediated regulation of gene expression provides plants with the molecular tools necessary to fine-tune their responses, ensuring survival and optimal growth under changing environmental conditions. Understanding these intricate signaling pathways enhances our grasp of plant biology and aids in developing strategies for improving crop resilience to environmental stressors.

\subsection{Physiological Roles}
ABA assumes diverse physiological roles pivotal for the survival and adaptation of plants. In biotic stress responses, ABA acts as a signaling molecule, triggering defense mechanisms against pathogens and herbivores. Moreover, ABA plays a significant role in fruit ripening, influencing the timing and coordination of this crucial developmental process. In morphological processes, ABA regulates root growth, impacting the plant's ability to explore soil for water and nutrients. Additionally, ABA is instrumental in stomatal development, influencing the opening and closing of pores crucial for gas exchange and water regulation. In instances of submergence, ABA orchestrates morphological changes in response to altered environmental conditions. These multifaceted physiological roles underscore the versatility of ABA in orchestrating adaptive responses to various stimuli, highlighting its importance in the intricate web of plant biology.

\subsection{Stress Avoidance and Adaptation}
Stress avoidance and adaptation are pivotal mechanisms in a plant's response to abiotic stress, with a particular emphasis on water-related challenges like dehydration. ABA emerges as a key player in this intricate process. As plants encounter abiotic stresses, ABA levels surge, serving as a signaling molecule that triggers a cascade of events aimed at bolstering the plant's adaptive capabilities. This hormone activates stress-responsive genes, orchestrating physiological changes that enhance the plant's resilience to adverse conditions. Notably, during water stresses, ABA induces stomatal closure, minimizing water loss through transpiration and aiding in stress avoidance. Understanding the nuanced interplay of ABA in stress avoidance and adaptation provides valuable insights for developing strategies to enhance crop resilience and agricultural sustainability.

\subsection{Biosynthesis}
ABA is a sesquiterpenoid hormone crucial in plant physiology, synthesized through the plastidic 2-C-methyl-D-erythritol 4-phosphate (MEP) pathway. Originating from isopentenyl diphosphate precursors, this pathway underlines ABA's biosynthesis within plant plastids. ABA plays multifaceted roles, notably responding to environmental stressors like drought and salinity. Its involvement in seed development, dormancy, stomatal regulation, and root development reflects its versatility in shaping plant growth and adaptation. ABA's signaling pathway encompasses receptors, kinases, and transcription factors, orchestrating intricate responses to external stimuli. Beyond plants, ABA has garnered attention for potential roles in non-plant organisms, underscoring its relevance in broader biological contexts. In sum, ABA's biosynthesis through the MEP pathway forms the molecular foundation for its diverse functions, illuminating its significance as a pivotal signaling molecule with far-reaching implications in the intricate tapestry of biological processes.

\section{Stimuli that increase the production of ABA}

There are various forms of stimuli that can trigger or halt the biosynthesis of ABA. These factors are not all stimuli. Some of them are regulatory procedures or natural events. Water Deficiency (Drought): When there is less water available, ABA levels frequently increase. Drought circumstances cause plants to synthesise more ABA, which helps the plant deal with water stress by encouraging stomatal closure and lowering transpiration loss. Osmotic Stress: Variations in osmotic potential have an impact on the synthesis of ABA. Osmotic stress caused by high salt or solute concentrations in the soil can raise ABA levels. Cold Stress: ABA synthesis can be stimulated by low temperatures. It contributes to the process of cold acclimation, enabling plants to withstand and adjust to cold stress. Pathogen attacks: ABA has a role in plants' defensive mechanisms against infections. ABA levels may rise in a plant's defensive mechanism in the event of a pathogen attack. Abscisic Acid Feedback Regulation: ABA itself can regulate its own synthesis. When ABA levels are elevated, it can use feedback to control its own production. Senescence: ABA is also involved in the regulation of senescence (ageing) in plants. As plants undergo natural senescence, ABA levels may increase. Given the characteristics of molecular communications we are going to inspect each of these stimuli from a signalling point of view. In each part a cascade of enzymes or hormones are discussed which eventually lead to biosynthesis of ABA. 

\subsection{Water Deficiency}

The Root of a plant is the first organ that senses the scarcity of water. This organ senses the drought and sends a signal to other parts of the plant. This signal is often in the form of calcium ions or reactive oxygen species. When this signal reaches some plant cells it triggers the activation of enzymes involved in the biosynthesis of ABA in the plastids of plant cells. The complete enzyme procedure is given the Table 1.

\begin{table}[ht]
    \centering
    \caption{Enzyme and Operation}
    \begin{tabular}{|p{6.4cm}|}
        \hline
        \textbf{Enzyme and Operation} \\
        \hline
        \textbf{DXS:} Catalyzes the conversion of pyruvate and D-glyceraldehyde 3-phosphate into 1-deoxy-D-xylulose 5-phosphate (DXP). \\
        \hline
        \textbf{DXR:} Acts on DXP to produce 2-C-methyl-D-erythritol 4-phosphate (MEP). \\
        \hline
        \textbf{HDR:} Catalyzes the conversion of MEP to 4-hydroxy-3-methylbut-2-en-1-yl diphosphate (HMBPP). \\
        \hline
        \textbf{CCD:} Enzymes cleave the carotenoid precursor, violaxanthin, to produce xanthoxin. \\
        \hline
        \textbf{NCED:} Catalyzes the oxidative cleavage of 9-cis-violaxanthin or 9-cis-neoxanthin to produce xanthoxin. \\
        \hline
        \textbf{ABA2:} Converts xanthoxin to abscisic aldehyde. \\
        \hline
        \textbf{AAO:} Oxidizes abscisic aldehyde to produce the biologically active form of ABA. \\
        \hline
        \textbf{SDR:} Converts xanthoxin to abscisic aldehyde in some plant species. \\
        \hline
    \end{tabular}
\end{table}

\subsection{Osmotic Stress}
This term refers to situations when the solute concentration is not the same inside and outside of the cell or organism. This difference of concentration influences the movement of water in one direction (Osmosis). The process is very similar to the way Water deficiency stimulates the production of ABA. This is expected since water deficiency causes osmotic stress. The chain of enzymes is similar to the last stimulus.
\subsection{Cold Stress}
The decrease in temperature triggers the change in permeability of the cell membranes. The rest of the effect of this stimulus can be visualised with a block diagram. 

\begin{figure}[ht]
    \centering
    \includegraphics[width=8.1cm]{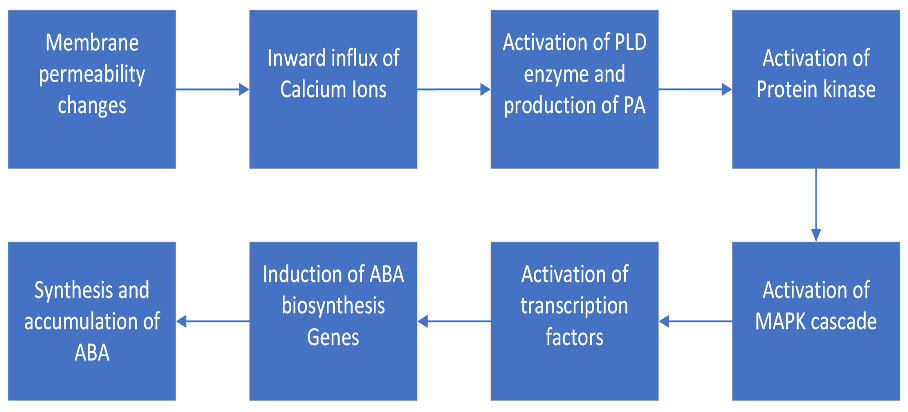}
    \caption{Cold Stress}
    \label{fig5}
\end{figure}

\subsection{Pathogen Attack}
\vspace{6pt}

It will first try to recognize the pathogen attack by recognizing the pathogen associated molecular patterns (PAMPs) using pattern recognition receptors (PRRs). After recognizing the pattern, the plant initiates a set of defensive responses known as PAMP triggered immunity (PTI), PTI involves various signalling including the activation of mitogen-activated protein kinase (MAPK) cascades. Activation of PTI results in intracellular calcium levels triggers the cascade discussed in the previous part. This way two different stimuli use the same cascade of enzymes. 

\subsection{Wounding}
\vspace{6pt}
When a plant undergoes physical damage, such as cell rupture caused by factors like wind or herbivores, it triggers a series of biochemical reactions. One consequence of cell rupture is the production of lipoxygenases, which leads to the synthesis of Oxilipins, including jasmonic acid (JA), an essential phytohormone. JA initiates a signaling pathway called the jasmonate signaling pathway, which is associated with the plant's response to injuries like wounding and herbivory. This signaling cascade involves the activation of various transcription factors. One important outcome of JA signaling is the transcriptional activation of genes involved in the biosynthesis of abscisic acid (ABA). As a result, these genes are transcribed and translated, leading to the synthesis of ABA near the ruptured cells. The accumulation of ABA at the site of damage induces various responses in the plant, such as stomatal closure to prevent further water loss, synthesis of protective compounds, and modulation of gene expression to enhance tolerance to environmental stress.

\begin{figure*}[!t]
    \centering
    \includegraphics{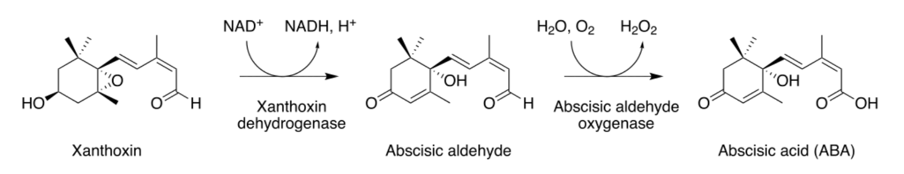}
    \caption{ABA Production Side}
    \label{fig6}
\end{figure*}

\subsection{Abscisic Acid Feedback Regulation}
\vspace{8pt}

High levels of ABA can trigger a process that will revert the changes and lower the level again. This mechanism prevents the accumulation of the hormone. The process involves several steps. 
ABA exists in various parts of the plant. In stress conditions there is a positive feedback loop that increases the amount of ABA as long as stress exists. But as soon as stress is mitigated (for example due to the closure of stomata the water is preserved) the biosynthesis of ABA is slowed.

The hormone itself is subject to degradation and catabolism which are basically molecules breaking down. One way it breaks down is with the enzyme ABA 8'-hydroxylases which breaks down ABA molecules into phaseic acid and dihydrophaseic acid. This process helps to decrease concentrations of ABA from plant tissues.Feedback inhibition of ABA biosynthesis: this mechanism refers to the way high levels of ABA in plant tissues downregulate the expression of certain genes that play a role in ABA biosynthesis. So it would not accumulate too much. Cross talk with other hormones in plants: ABA molecules can in fact interact with other hormones such as gibberllins, cytokinins, auxins and etc. these interaction can impact the overall hormonal balance of plants with regards to ABA. 

\subsection{Ageing (Senescence)}
\vspace{8pt}
Senescence is one of the key stages of plants which is programmed into the cell. Senescence and ABA levels are related to each other in other words, ABA levels play a role in Senescence and as Senescence progresses it will affect ABA levels. The accumulation of ABA in plant tissues is part of the senescence process. It causes the plant to move nutrients from the ageing parts to the parts with more life left. During Senescence the expression of genes that play a role in biosynthesis of ABA is increased. A specific enzyme known as 9-cis-epoxycarotenoid dioxygenase plays a role in this process. (“Double NCED isozymes control ABA biosynthesis for ripening and senescent regulation in peach fruits”) Also there exists interactions with other hormones in the context of Senescence such as ethylene and jasmonic acid. These interactions control the progression of Senescence to a certain degree. It should be noted that some environmental effects such as stress can affect the relationship between ABA and senescence. For example stress can increase the speed of senescence and therefore increase ABA levels. In conclusion, there are multiple paths that lead to generation or degradation of ABA. These processes sometimes overlap in various ways. It is difficult to separate these signalling pathways in the manner that we did here. But it gives a clear big picture.

\section{Production Site of ABA}
\vspace{8pt}
The Abscisic acid (ABA) is a crucial isoprenoid plant hormone synthesized through the plastidal 2-C-methyl-D-erythritol-4-phosphate (MEP) pathway, distinguishing it from sesquiterpenes, which derive from farnesyl diphosphate (FDP) in the mevalonic acid pathway. The C15 backbone of ABA originates from the cleavage of C40 carotenoids within the MEP pathway, with zeaxanthin serving as the initial committed ABA precursor. Through a sequence of enzyme-catalyzed epoxidations and isomerizations involving violaxanthin, and subsequent dioxygenation-induced cleavage of the C40 carotenoid, the proximal ABA precursor, xanthoxin, is formed. This intermediary is then further oxidized to ABA via abscisic aldehyde. Notably, the development of Abamine marks a significant milestone, representing the first specific inhibitor of ABA biosynthesis. Its synthesis and patenting provide a means to effectively regulate endogenous ABA levels, offering valuable insights into the intricate process. 

For instance the Arabidopsis thaliana plant exhibits distinct pathways for both the biosynthesis and catabolism of abscisic acid (ABA). The progression of de novo ABA synthesis from zeaxanthin to xanthoxin takes place within plastids. Subsequently, xanthoxin translocates from plastids to the cytoplasm, where it undergoes conversion into ABA. In the catabolic pathways, ABA undergoes inactivation through oxidation or conjugation processes. Enzymes facilitate the hydroxylation and hydrolysis of ABA-glucosyl ester (ABA-GE), leading to the formation of phaseic acid (PA) and dihydrophaseic acid (DPA). The depicted pathways include both confirmed pathways, represented by solid lines, and postulated pathways, illustrated by broken lines. Noteworthy cellular compartments include the endoplasmic reticulum (ER). This comprehensive view encapsulates the intricate interplay between ABA biosynthesis and catabolism in Arabidopsis, shedding light on the regulatory mechanisms governing ABA levels and its derivatives.

\begin{figure}[ht]
    \centering
    \includegraphics[width=0.95\linewidth]{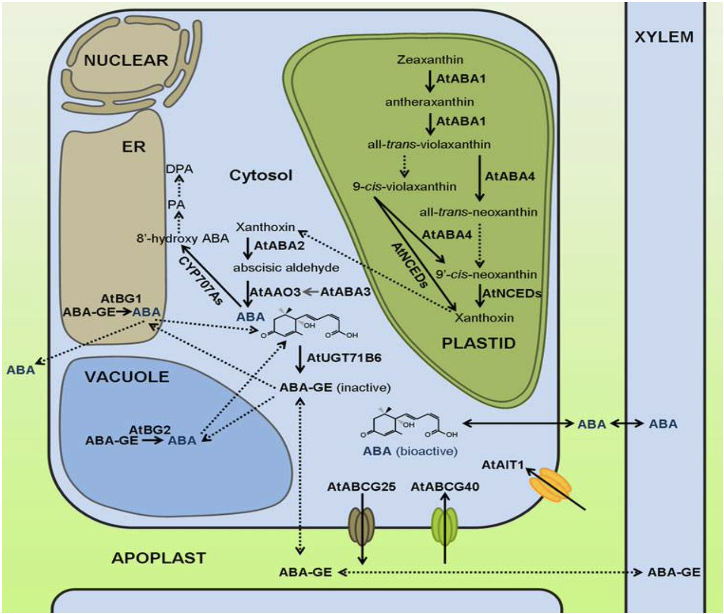}
    \caption{Pathway of ABA}
    \label{fig8}
\end{figure}

\subsection{Biosynthesis Pathway - Unraveling the Genetic Machinery}
The primary route for abscisic acid (ABA) biosynthesis involves the transformation of carotenoids, orchestrated by specific genes. Notably, AtABA1 codes for Zeaxanthin epoxidase, a catalyst for the conversion of zeaxanthin to all-trans-violaxanthin, while AtABA4 facilitates the transition from violaxanthin to neoxanthin. The pivotal cleavage step, leading to the formation of xanthoxin, is mediated by nine-cis-epoxycarotenoid dioxygenase (NCED) enzymes Intriguingly, the final two steps, involving xanthoxin conversion to ABA, occur in the cytosol, necessitating transport from plastids. AtABA2 and abscisic aldehyde oxidase play crucial roles in these final steps, underscoring the intricate spatial orchestration of ABA biosynthesis. Mutant studies in Arabidopsis, tomato, and maize have been instrumental in elucidating the biochemical intricacies at each stage, advancing our understanding of de novo ABA production \cite{b13}.

\begin{figure}[ht]
    \centering
    \includegraphics[width=0.95\linewidth]{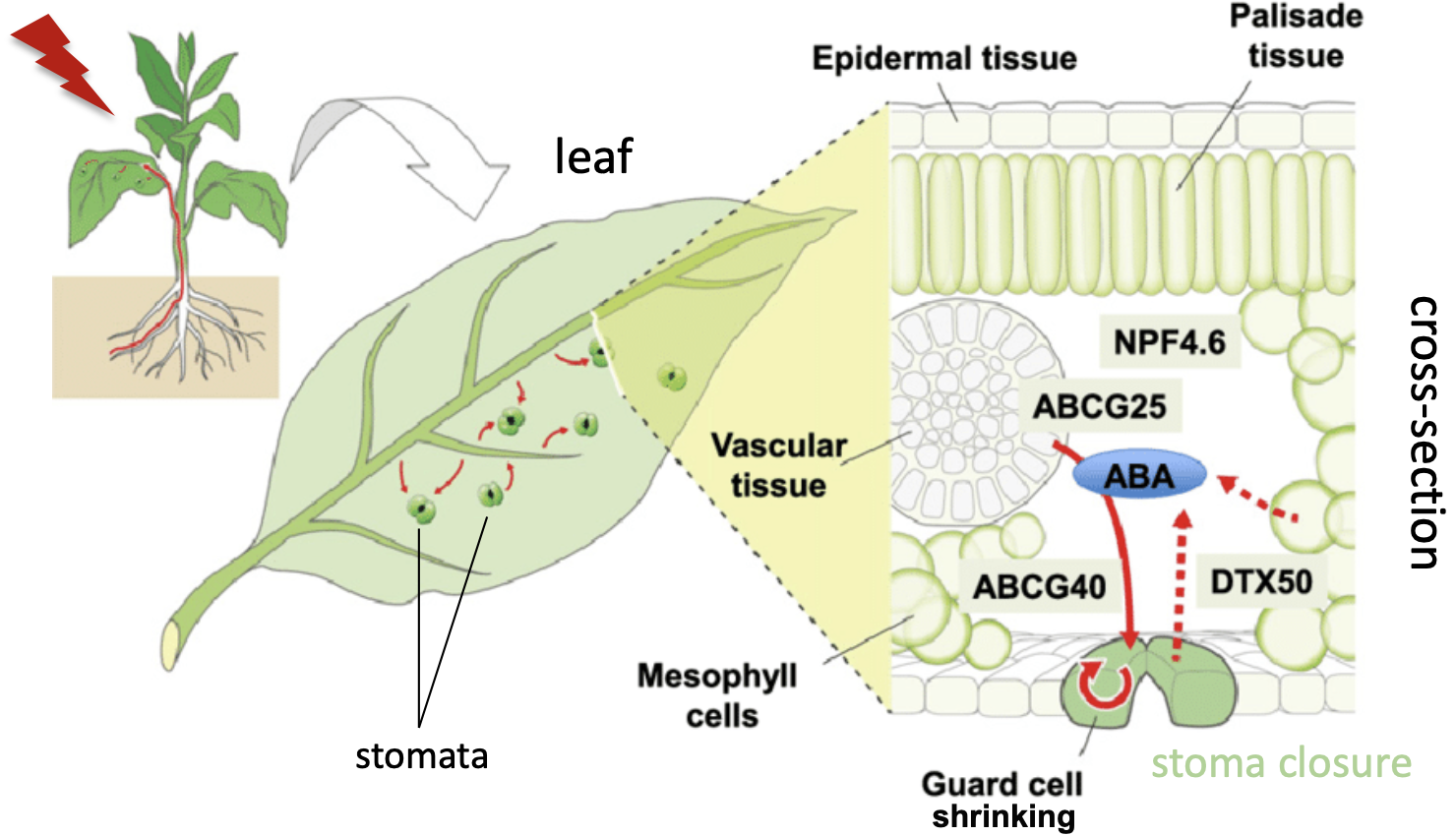}
    \caption{ABA Synthesis}
    \label{fig9}
\end{figure}

\subsection{Intriguing Pathway: ABA Synthesis from ABA-GE}

An alternative ABA biosynthesis pathway involves the hydrolysis of ABA glucose esters (ABA-GE) by $\beta$-glucosidase homologs, namely AtBG1 and AtBG2. These enzymes, located in the endoplasmic reticulum and vacuole, respectively, facilitate a rapid and direct conversion of ABA-GE to ABA. This single-step process contrasts with the lengthy de novo pathway, particularly advantageous under abiotic stress conditions. Loss-of-function and gain-of-function mutants, such as the severe ABA-deficient phenotype in the atbg1 mutant and the mild phenotype in the atbg2 mutant, suggest potential divergent physiological roles for ABA produced by these homologs. This raises intriguing questions about the distinctive functions of ABA generated in different cellular compartments and the underlying regulatory mechanisms \cite{b14}.

\subsection{Physiological Implications ABA and Cellular Homeostasis}

The emerging understanding of ABA biosynthesis in distinct cellular compartments poses fascinating questions regarding its physiological implications. The coordination of ABA production pathways in different compartments plays a pivotal role in maintaining cellular ABA levels, particularly in response to changing environmental conditions. As evidenced by the diverse phenotypes observed in mutants, the intricate interplay between AtBG1, AtBG2, and other ABA biosynthetic genes contributes to the dynamic regulation of ABA levels. Exploring how ABA produced in different compartments influences cellular homeostasis opens avenues for comprehending the multifaceted roles of ABA in plant adaptation and stress responses. 

\section{Transport of ABA in the plant}
\vspace{6pt}
\subsection{Transport in the Plant: A Crucial Role in Plant Physiology}
\vspace{6pt}
Abscisic acid (ABA), a pivotal plant hormone, intricately orchestrates a spectrum of physiological processes, notably responding to environmental stressors such as drought, salinity, and cold. Its regulatory reach extends to crucial aspects like plant water status, stomatal closure, seed dormancy, and nuanced responses to environmental cues. Central to this intricate orchestration is the transport of ABA within plants, a process predominantly facilitated by the xylem and phloem the primary vascular tissues responsible for the dynamic movement of water, nutrients, and signalling molecules.

The xylem emerges as the primary conduit for ABA transport, playing a vital role in long-distance signalling during stress, particularly drought. This process involves the synthesis of ABA in the roots and its subsequent journey through the xylem to influence shoots, ultimately impacting stomatal conductance and other physiological responses. Complementing the xylem's role, the phloem, designed for the transport of organic compounds, including hormones, contributes to the comprehensive distribution of ABA within the plant.

Understanding the intricacies of ABA transport dynamics unravels the mechanisms governing plant adaptation to environmental challenges. From symplastic and apoplastic pathways to computational models and scientific insights, a holistic comprehension of ABA transport positions us to develop strategies enhancing crop resilience and productivity amidst evolving environmental conditions.
\vspace{6pt}
\subsection{Long Distance Signalling in Drought Stress}
\vspace{6pt}
The xylem stands as the primary conduit for the transport of abscisic acid (ABA) within plants, particularly wielding significance in the context of drought stress. Functioning primarily to transport water and minerals from roots to shoots, the xylem emerges as a crucial pathway for ABA movement. Under the duress of drought stress, ABA synthesised in the roots embarks on a journey through the xylem to the shoots. This translocation plays a pivotal role in influencing stomatal conductance and eliciting physiological responses crucial for the plant's adaptation to water scarcity.

The pH of the xylem sap assumes a significant role in the dynamics of ABA transport and accumulation within the plant. This nuanced interaction between ABA and the pH of xylem sap adds a layer of complexity to the regulatory mechanisms associated with ABA signalling. As ABA traverses the xylem, the pH conditions intricately modulate its transport efficiency and subsequent accumulation in various plant tissues. This multifaceted interplay within the xylem underscores its paramount role as the chief conduit for ABA, unravelling a key aspect of plant response mechanisms to the challenging conditions imposed by drought stress. 

\begin{figure}[ht]
    \centering
    \includegraphics[width=0.95\linewidth]{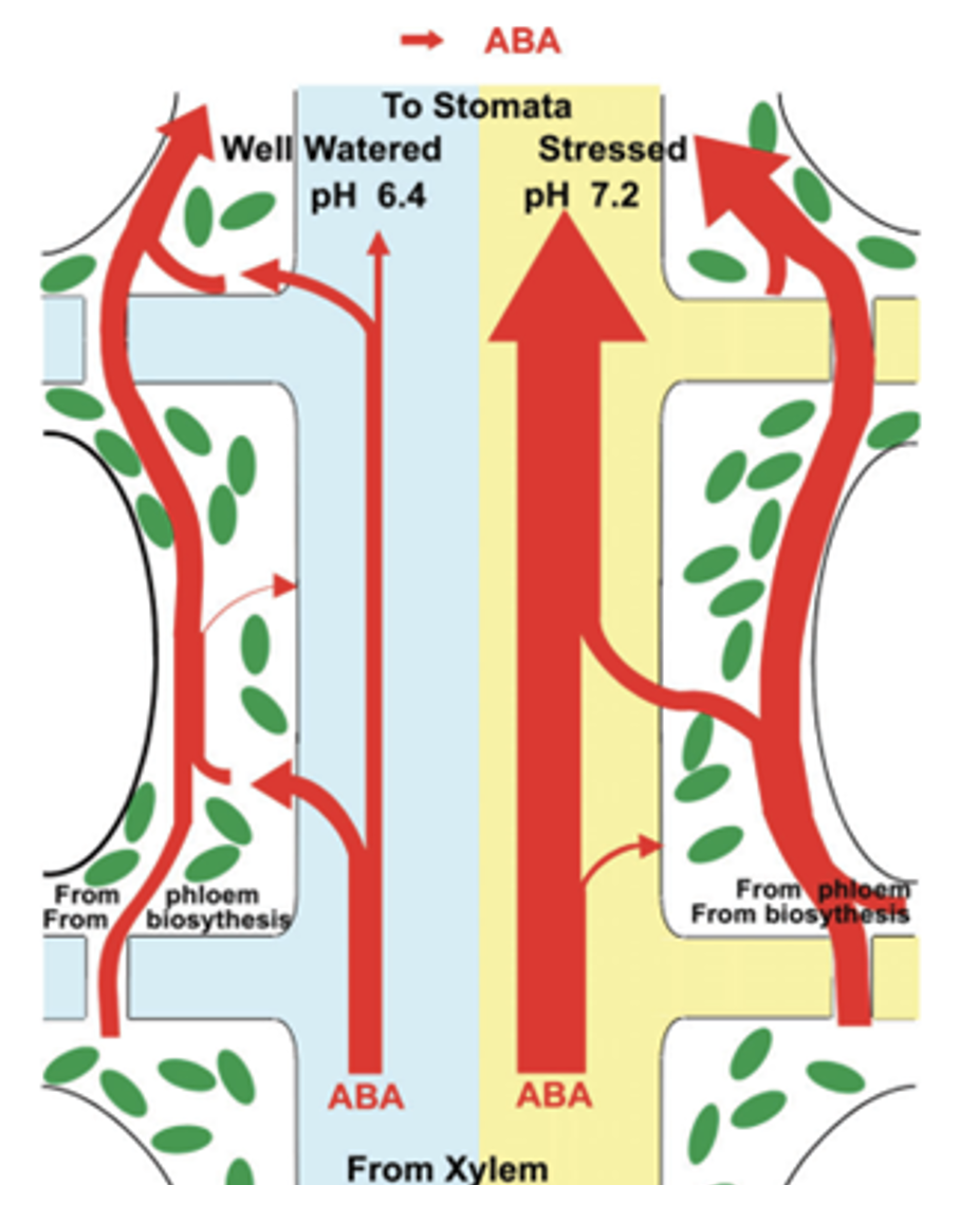}
    \caption{ABA Transport}
    \label{fig10}
\end{figure}

\begin{figure}[ht]
    \centering
    \includegraphics[width=0.95\linewidth]{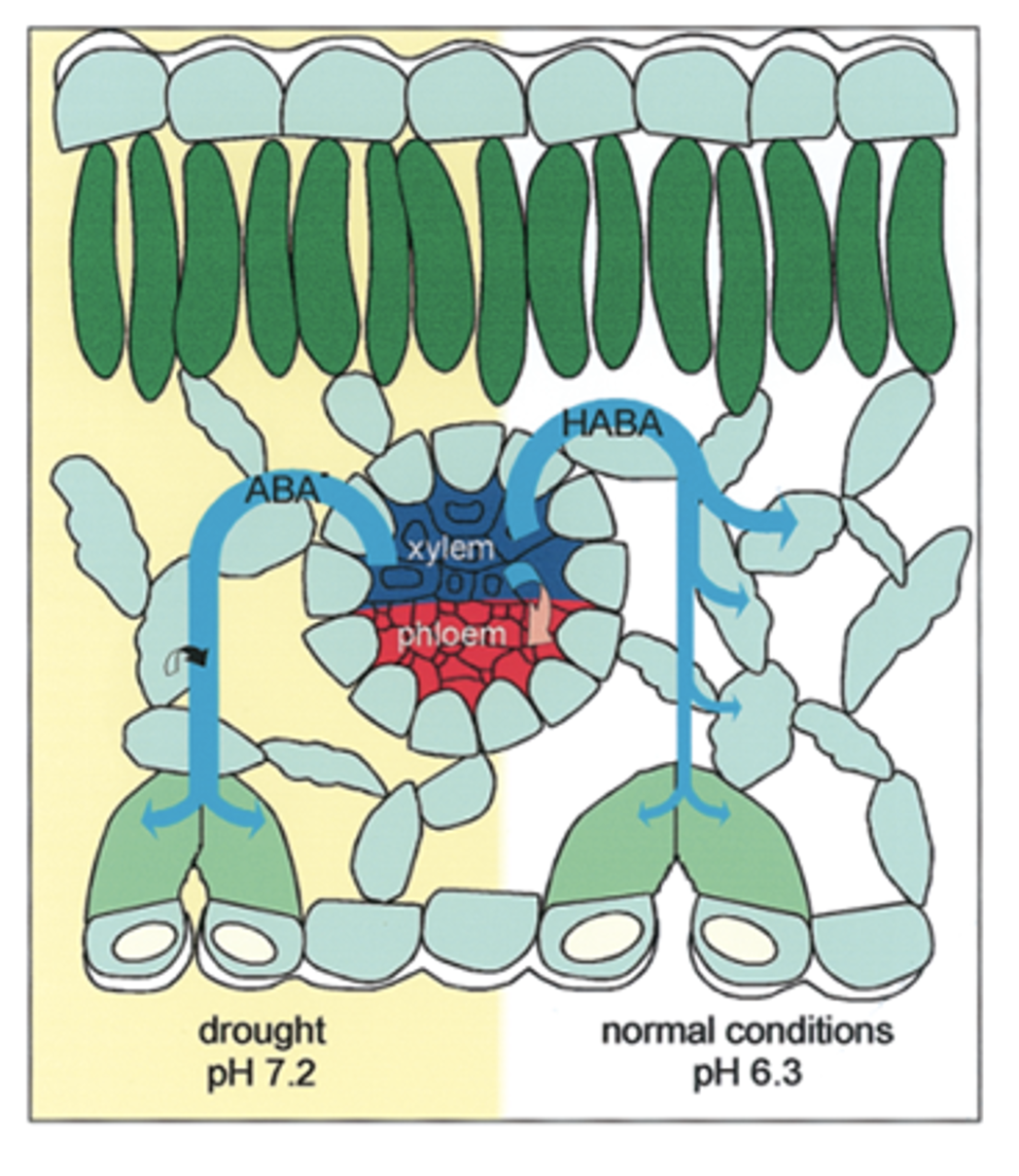}
    \caption{ABA Transport}
    \label{fig11}
\end{figure}

\subsection{Coordinated Responses to Stress}

Abscisic acid (ABA) transport within plants unfolds through two pivotal pathways: symplastic and apoplastic, each contributing to coordinated responses under stress. Symplastic transport involves the movement of ABA through interconnected plant cell cytoplasm, allowing for direct transfer between cells. This intricate pathway facilitates swift and coordinated stress responses across various parts of the plant. Apoplastic transport, on the other hand, takes place through cell walls and intercellular spaces, shaping ABA's mobility and availability within the plant. Symplastic transport establishes a direct cytoplasmic route, fostering quick and synchronised communication between plant cells. This interconnectedness enables the plant to orchestrate cohesive responses to environmental stressors, particularly drought, salinity, and cold. Apoplastic transport, characterised by movement through cell walls and intercellular spaces, influences the accessibility and distribution of it. This pathway contributes to the strategic deployment of it in different plant tissues, ensuring a dynamic and responsive approach to stress conditions. In essence, the symplastic and apoplastic transport pathways illustrate the versatility of ABA movement within the plant, allowing for nuanced and coordinated reactions to environmental challenges. The interplay between these pathways underscores the sophistication of plant adaptation mechanisms, providing a foundation for further exploration into the intricacies of ABA-mediated stress responses.

\subsection{Computational Models}

In the realm of plant physiology, computational models stand as invaluable tools, propelling our comprehension of abscisic acid (ABA) transport dynamics to new heights. These sophisticated models delve into the intricate processes of ABA synthesis, degradation, and transport through both xylem and phloem pathways. Through meticulous simulations, they unravel the complex interplay of factors governing ABA movement within plants. These models extend their utility beyond mere simulations, emerging as predictive instruments for anticipating plant responses to diverse environmental conditions. One of the paramount contributions of computational models lies in elucidating ABA's pivotal role in regulating water use efficiency, particularly under stressful conditions. By simulating ABA dynamics, these models offer insights into how the hormone influences physiological responses, ultimately shaping a plant's ability to thrive in the face of environmental challenges. Their predictive power becomes especially crucial in deciphering the nuanced mechanisms that plants employ to adapt to varying stressors. In essence, computational models serve as illuminating windows into the world of ABA transport, unravelling its complexities and providing a foundation for understanding how plants navigate and respond to their ever-changing surroundings. As we harness the capabilities of these models, we gain not only insights into the intricacies of plant physiology but also a pathway towards developing strategies to enhance crop resilience and optimise water use efficiency in agriculture.

\subsection{ABA Transport Mechanisms Unveiled}

Scientific papers provide intricate insights into the mechanisms governing the transport of abscisic acid (ABA), unravelling nuanced details crucial for understanding plant responses to stress. These papers particularly underscore the pivotal role of the xylem in facilitating long-distance signalling during stress. Lateral flows of ABA, the chemical characteristics of Casparian bands, and the mechanisms governing the release of ABA glucose ester (ABA-GE) are meticulously elucidated. Of particular significance is the revelation that permeability coefficients of stem parenchyma plasma membranes play a central role in the redistribution of ABA between stem parenchyma and xylem. This intricate interplay significantly contributes to maintaining ABA homeostasis within the plant.

\begin{figure}[ht]
    \centering
    \includegraphics[width=8cm, height=10cm]{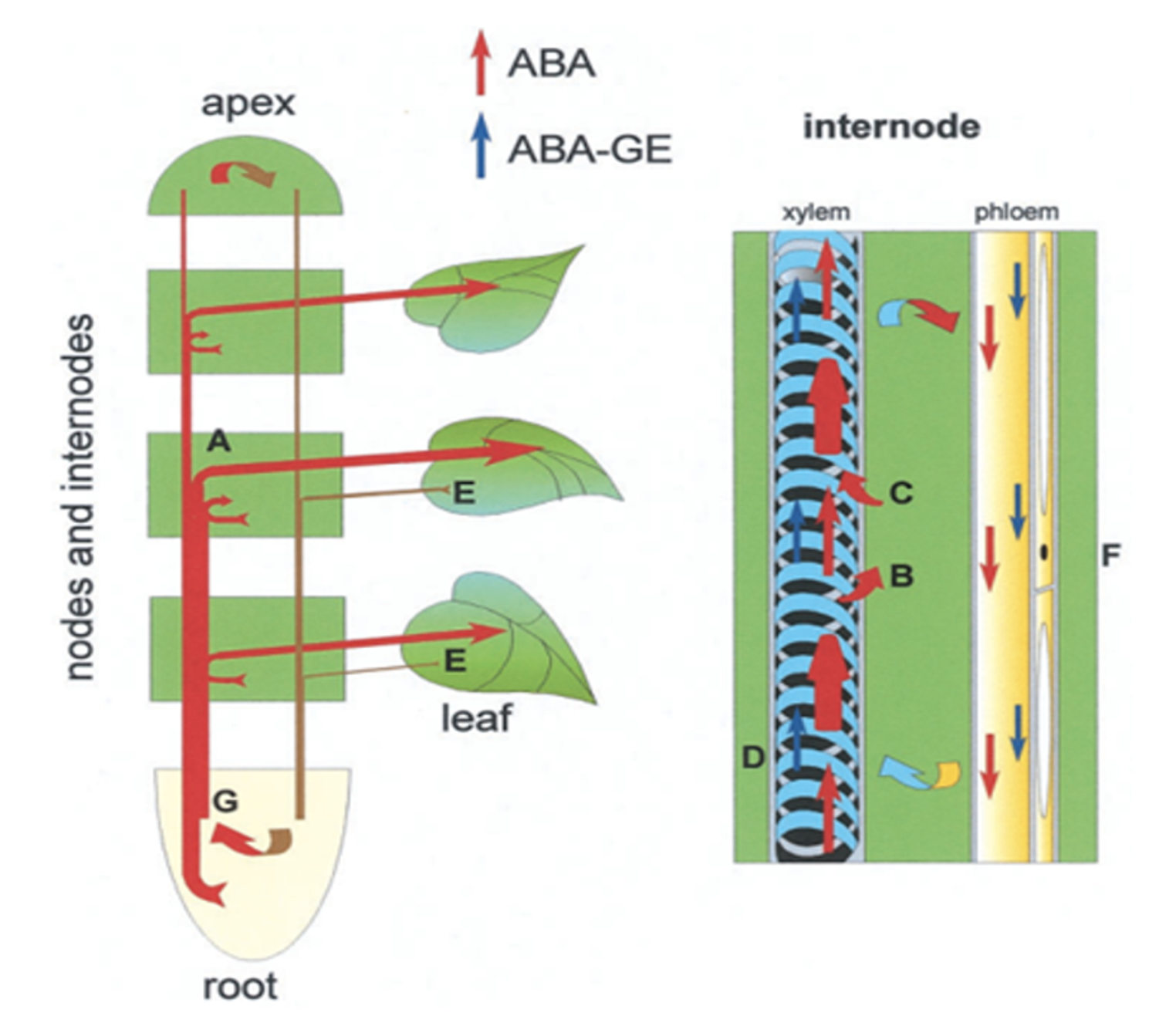}
    \caption{ABA Transport Mechanisms }
    \label{fig12}
\end{figure}

\begin{figure}[ht]
    \centering
    \includegraphics[width=8cm, height=10cm]{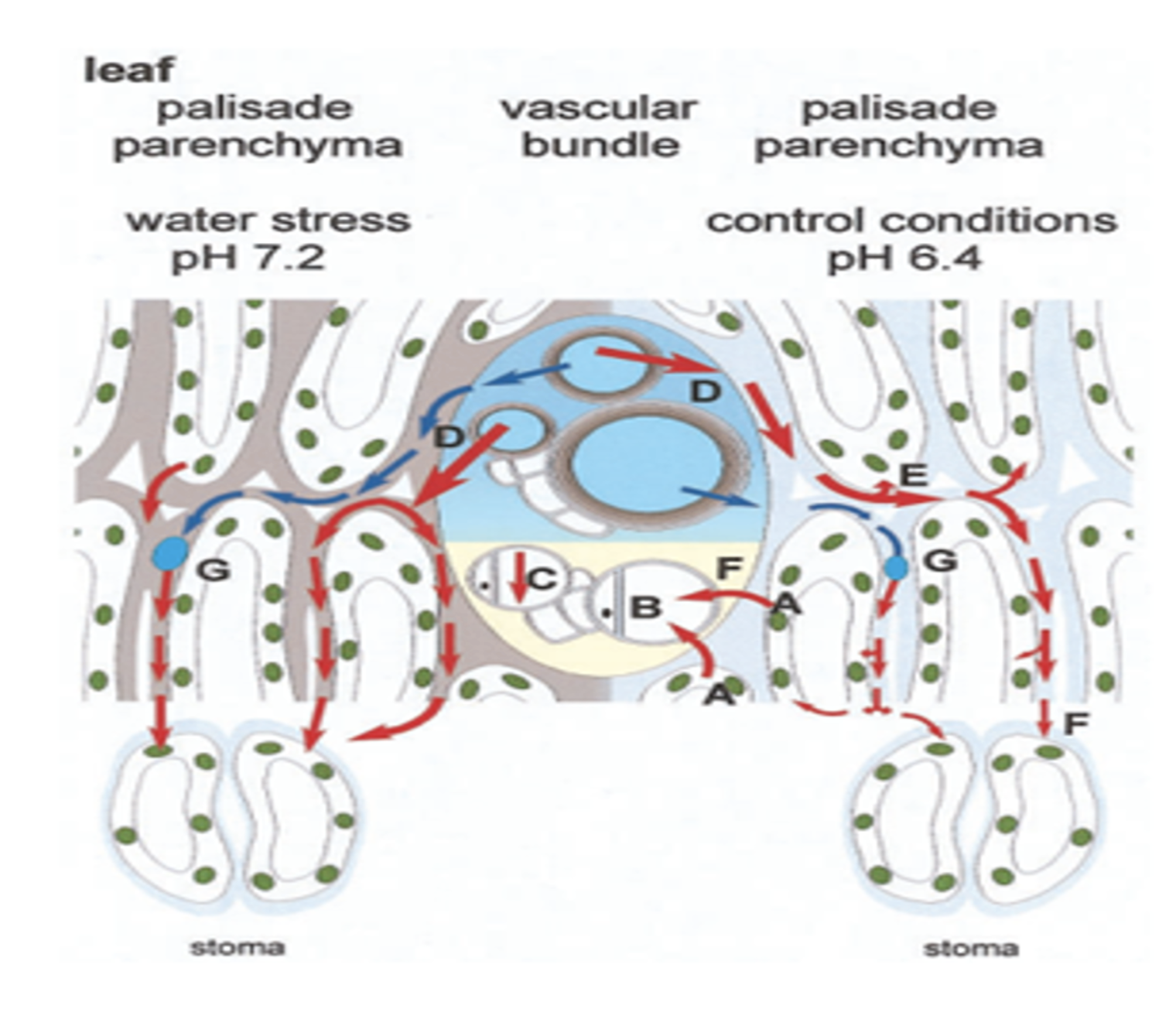}
    \caption{ABA Transport Mechanisms }
    \label{fig13}
\end{figure}

The detailed examination of ABA transport mechanisms sheds light on the complexity of its journey, highlighting key factors that regulate its distribution and signalling functions. Understanding the dynamics of lateral ABA flows and the impact of Casparian bands' chemical properties provides a comprehensive picture of ABA transport. Moreover, the insight into ABA-GE release mechanisms adds another layer of complexity to the regulatory processes governing ABA homeostasis. Altogether, these revelations from scientific papers deepen our grasp of the intricate interplay within plant physiology, offering valuable knowledge for devising strategies to enhance plant resilience under stress conditions.

\begin{figure}[ht]
    \centering
    \includegraphics[width=8cm, height=10cm]{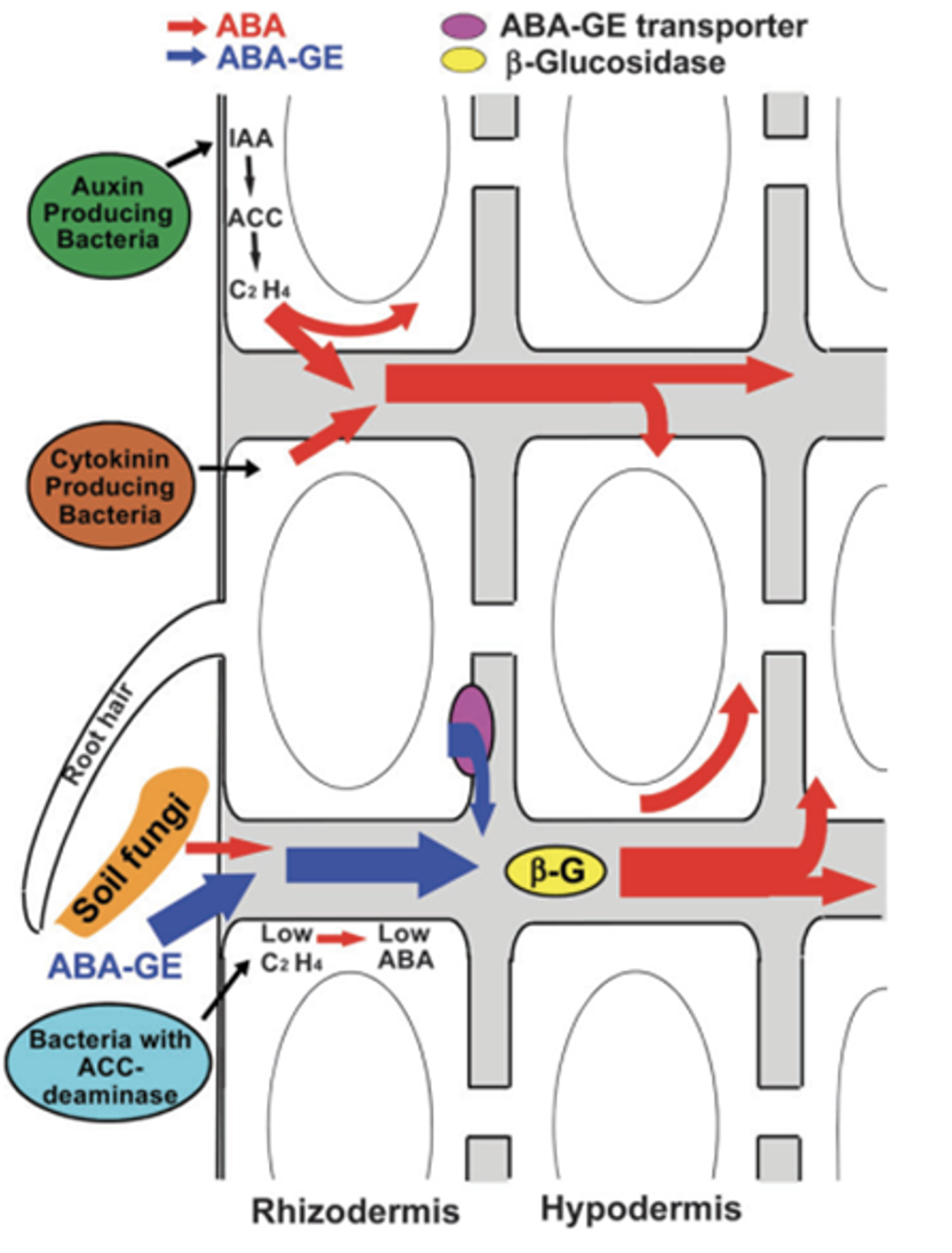}
    \caption{ABA Transport Mechanisms }
    \label{fig14}
\end{figure}

\begin{figure}[ht]
    \centering
    \includegraphics[width=8cm, height=10cm]{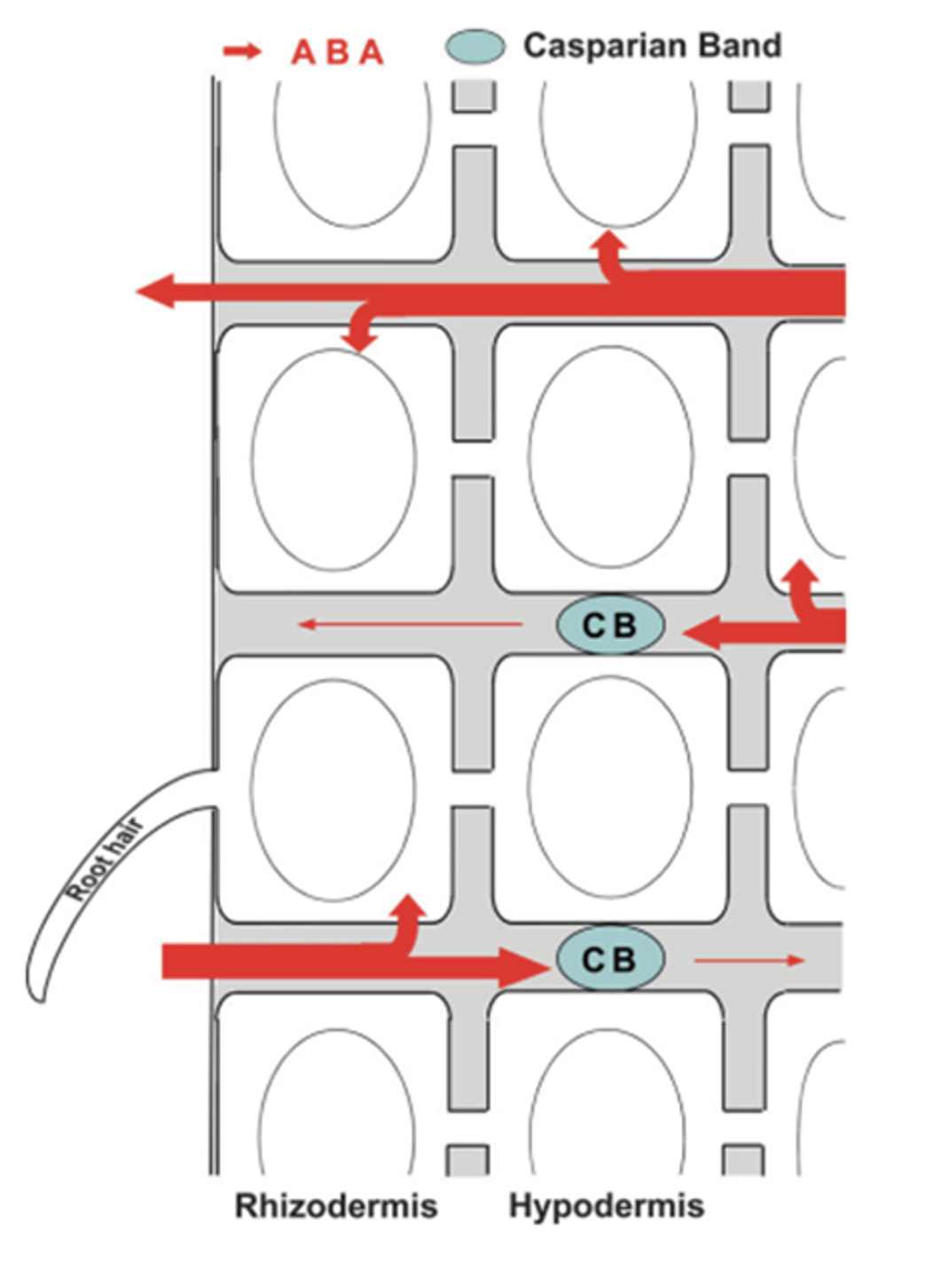}
    \caption{ABA Transport Mechanisms }
    \label{fig15}
\end{figure}

\subsection{Conclusion: Key to Plant Adaptation and Crop Improvement}

In summary, the transport of abscisic acid (ABA) within plants stands out as a pivotal factor influencing plant adaptation to environmental challenges. Delving into the symplastic and apoplastic pathways, acknowledging the contribution of computational models, and assimilating intricate insights from scientific papers significantly enhances our grasp of ABA dynamics in plant physiology. This profound understanding becomes indispensable for formulating strategies aimed at bolstering crop resilience and productivity amid the ever-changing environmental conditions. The symplastic and apoplastic transport pathways unravel the nuanced ways in which ABA moves within plant cells and through intercellular spaces, coordinating responses to stress. The integration of computational models provides a predictive framework, offering valuable foresight into ABA distribution and its consequences under diverse environmental scenarios. Scientific papers, with their detailed examinations of ABA transport mechanisms, contribute essential nuances, emphasising the role of the xylem, the impact of external factors, and the dynamic interplay of ABA forms. Armed with this comprehensive knowledge, scientists and agriculturists can design targeted interventions to optimise ABA-mediated responses, thereby fortifying crops against environmental adversities. Ultimately, the exploration of ABA transport dynamics unveils a critical nexus for unlocking the potential of plant adaptation and cultivating strategies that elevate agricultural sustainability in the face of an ever-evolving environment.

\section{Experimental quantification of ABA in plant}
\subsection{What is (GLC-EC)}
Liquid Chromatography with Electrochemical Detection (GLC-EC) represents a potent analytical method. Used for example Abscisic Acid (ABA), an essential plant hormone participating in stress responses and diverse physiological functions. GLC works in separating mixtures quickly and Gas- specially with EC, it provides highly responsive detection capabilities for ABA Electrochemical detection, by virtue of allowing distinctive electrochemical reactions, contributes to heightened specificity as ABA is uniquely identifiable through this process. Through the creation of calibration curves using known concentrations of ABA standards, GLC-EC facilitates precise and dependable quantification of ABA levels in plant samples. This comprehensive method proves essential in progressing research in plant biology, fostering a more profound comprehension of ABA's role in plant stress responses.

\subsection{What is HPLC with a UV detector}
Analytical High-Performance Liquid Chromatography (HPLC) with a UV detector is a main dependent method for quantifying Abscisic Acid (ABA), an essential plant hormone participating in diverse physiological processes, including responses to stress. In this analytical method, a liquid sample containing ABA is introduced into a high-pressure liquid chromatography, where it undergoes separation according to its chemical characteristics. 

\

\

The UV detector is designed to specifically focus on ABA molecules, enabling their identification and measurement. UV detection proves highly effective for compounds like ABA that absorb UV light, offering a sensitive and selective analysis method. Through the comparison of the sample's UV absorption with established ABA standards, the HPLC-UV configuration allows for precise determination of ABA concentrations in plant samples. This significantly contributes to research in plant physiology and stress signalling pathways.

\begin{figure}[!ht]
    \centering
    \includegraphics[width=8.2cm, height=4cm]{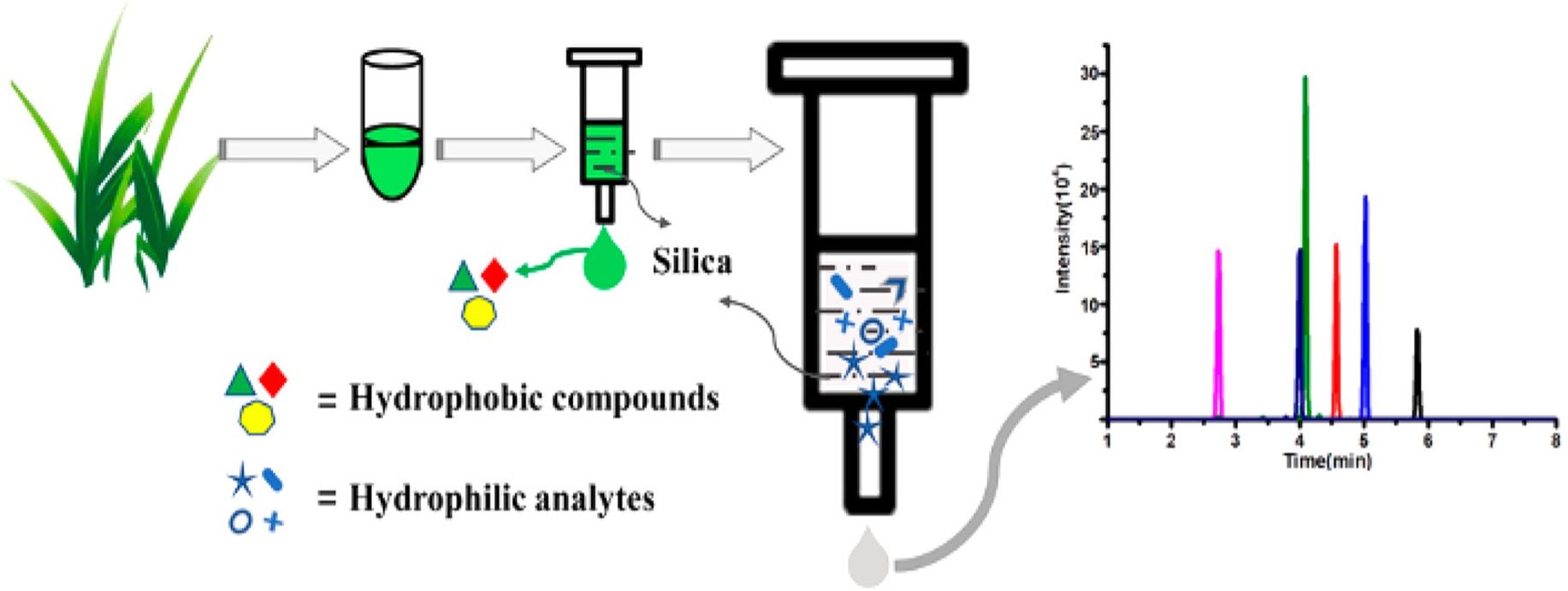}
    \caption{HPLC with a UV detector}
    \label{fig16}
\end{figure}

\

\

\

\subsection{GLC-EC VS HPLC}
In chromatographic techniques, the choice of mobile phase, separation principles, operating conditions, applications, and equipment distinguishes High-Performance Liquid Chromatography (HPLC) and Gas Chromatography (GC). HPLC employs a liquid mobile phase selected based on sample characteristics, emphasising factors like polarity and solubility. Separation in HPLC is influenced by compound interactions with both mobile and stationary phases. The method operates at room temperature, and runs typically last 10-60 minutes. On the other hand, GC uses an inert gas as the mobile phase, selected based on the detection method, and separation depends on compound volatility, with more volatile molecules moving faster. GC requires higher temperatures (150°C), but it's runs are quicker, often in minutes or seconds. HPLC is suitable for analysing soluble compounds, commonly used in food, water, and polymer analysis, while GC is ideal for volatile mixtures, making it suitable for drugs, toxins, and air samples. Equipment-wise, HPLC utilises shorter, wider columns and requires expensive solvents, while GC features long, thin columns and is more cost-effective, using gas containers and carrier gas. Overall, these differences make each technique more suitable for specific analytical needs

\begin{figure}[ht]
    \centering
    \includegraphics[width=8cm, height=5cm]{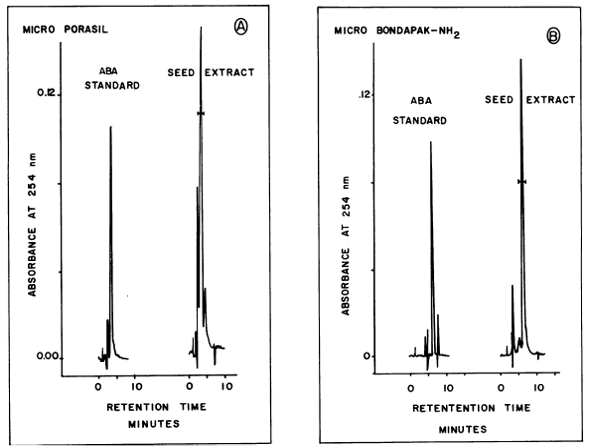}
    \caption{Analytical Quantification by GLC-EC}
    \label{fig17}
\end{figure}

\subsection{Experiment on quantifying the ABA in Soybean}
\subsubsection{Analytical Quantification by GLC-EC}
ABA quantification on soybean can be conducted by GLC-EC of plant extracts after a previous procedure of HPLC preparation. Electrons that capture detectors for molecules with high electron affinity are selected, and a large number of quantities of organic compounds other than ABA were observed also in the sample. With the detector used in this study, the lower limit of sensitivity was 10 pg ABA/injection. Some of these organic compounds did not dissolve well in the 10\% (v/v) methanol in either solution used for methylation. Methanol was added initially to dissolve to ascertain if ABA was trapped within inadequately dissolved elements, hindering the full methylation of all ABA molecules. Absorbance (254 nm) of standard ABA (500 ng) g±Bondapak-NH2 (B). Ether was added to achieve the desired concentration, and subsequently, the sample was introduced. The final yield of methylated ABA remained largely unaffected by concentrations of methanol ranging from 10\% to 50\% in ether.
Then, the sample it died with N2 stream at temperature 30 C and reconstituted in a After methylation, the sample was taken to dryness with N2 stream at 30 C and then reconstituted in a 
Appropriate solvent for gas-liquid chromatography (GLC). Hexane, ethyl acetate, pyridine, and acetonitrile are commonly used solvents for this purpose. For picking the solvent, Firstly, ABA must be completely soluble, Secondly, solvent peak tailing must be minimal; Lastly The duration for the detector signal to revert to its baseline after sample injections should be kept to a minimum. Hexane, ethyl acetate, and acetonitrile gave very little solvent peak tailing before the time of ABA elution, but did not completely dissolve the ABA sample. Pyridine completely dissolved the sample, but had a large solvent peak tailing.

A 10\% (v/v) solution of pyridine in hexane was the final choice. The sample treated with methylation was dissolved in 50 microliters of pyridine, followed by the addition of 450 microliters of hexane. The solubility in pyridine demonstrated excellence, and acceptable solvent peak tailing was observed. However, the solvent choice may vary depending on the plant material, the GLC column packing, and the electron capture detector used. Therefore, it is essential to carefully explore different solvents for achieving efficient and reproducible quantification of ABA through GLC-EC. Utilising multiple ion detection on GC-MS for the primary ABA fragments (91- 190, m/e) 7 confirmed the presence of ABA in the seed sample. A distinctive fragmentation pattern specific to ABA was only observed at the retention time corresponding to the ABA standard.

\subsection{Analytical Quantification by HPLC.}
ABA can also be quantified in another way using HPLC (High-performance liquid chromatography), HPLC in suitable for large injection volumes, that allow quantification a big sample of the total portion that is injected. However, Analytical HPLC quantification necessitates a sample dilution 40 times less than GLC-EC, indicating a potential for greater precision with the former method. Among the three HPLC columns examined In previous studied steps, Both Porasil and Bondapak-NH2 were determined to be appropriate for the quantification of ABA in a soybean extract using the preparative HPLC method previously outlined. Porasil enables the quantification of ABA using 15\% (v/v) acetonitrile in chloroform, acidified with 0.17 N acetic acid at a flow rate of 2 ml/min. The lower limit of detection was established at 500 pg ABA per injection, exhibiting linear quantification up to 20 µg when employing a UV detector (254 nm). Similar ABA quantification results were obtained with sample injections of either 20 µl or 200 µl. Bondapak-NH2 allowed for the quantification of ABA using 45\% (v/v) acetonitrile in chloroform, acidified with 0.17 N acetic acid at a flow rate of 2 ml/min. The lower limit of detection and the linearity of concentration were similar to those observed with Porasil.

Specific Bondapak, while separating ABA and t-ABA standards with a solution of 45\% (v/v) methanol in 0.2 N acetic acid at a flow rate of 4.5 ml/min, faced challenges in quantifying ABA from a plant extract following preparative HPLC due to interfering UV-absorbing compounds. This issue might be addressed by altering the organic solvent and/or pH used to elute the column or by derivatizing the ABA.Filtering had no impact on ABA concentration, and 95\% of the radioactivity was recovered in the ABA-containing fraction from each column. Analytical HPLC with Porasil and Bondapak-NH2 columns has a lower limit of detection of 500 pg ABA/injection, whereas GLC-EC has a lower limit of 10 pg/injection. Elution times and turnaround times for ABA were comparable in both analytical HPLC and GLC-EC. Unlike GLC-EC, analytical HPLC does not necessitate sample derivatization and facilitates simpler sample collection post-quantification. However, a notable drawback in comparison to GLC-EC is the UV detector's lower selectivity, presenting challenges when ABA levels are akin to those of interfering compounds. For such samples, additional purification before analytical HPLC may be required. This limitation could be overcome by employing more selective detectors, reducing the need for extensive sample purification in complex samples.

Moreover, laboratories engaging in substantial work with ABA, and potentially other plant hormones, may find HPLC advantageous, providing both preparative clean-up and analytical quantification using the same instrumentation. It is crucial to emphasise that utilising HPLC for the identification and quantification of plant hormones may be precarious unless the peaks being identified are meticulously characterised as pure, preferably through bioassay or GC-MS.

\section{Modelling ABA transport in plant}

\subsection{Mathematical Modelling}

\subsubsection{Diffusion Theorem}

The diffusion theorem elucidates the process by which particles disperse through a medium over time. Fundamentally, it describes the movement of molecules from regions of high concentration to regions of low concentration. Mathematically, this phenomenon is encapsulated by Fick's first law, which states that the flux of diffusing particles (\( J \)) is proportional to the concentration gradient (\( \frac{\partial c}{\partial x} \)), with the diffusion constant (\( D \)) as the proportionality constant. In one dimension, this relationship is given by:

\[
J = -D \frac{\partial c}{\partial x}
\]

Moreover, the continuity equation, a manifestation of the conservation of matter, characterizes how the concentration (\( c \)) evolves over time and space. It is represented by the diffusion equation:

\[
\frac{\partial c}{\partial t} = D \frac{\partial^2 c}{\partial x^2}
\]

This equation showcases how the concentration at a particular point changes with respect to both time and space. Additionally, in three dimensions, the diffusion equation extends to include radial dependence (\( r \)) as follows:

\[
\frac{\partial c}{\partial t} = D \frac{\partial^2 c}{\partial r^2}
\]

This equation provides insights into how concentration evolves in a three-dimensional space over time due to diffusion. Overall, the diffusion theorem, encapsulated by Fick's laws and the diffusion equation, serves as a cornerstone in understanding the dynamics of molecular movement in various physical and biological systems.

\subsubsection{Brownian Motion Explanation}

Brownian motion, a fundamental concept in statistical physics, describes the seemingly random movement of particles suspended in a fluid medium. Initially observed by Robert Brown in 1827, this phenomenon arises due to the incessant collisions between the suspended particles and the surrounding fluid molecules. Mathematically, Brownian motion is often modeled using the Langevin equation, which characterizes the motion of a particle subjected to both deterministic and random forces. The Langevin equation, given by:

\[
m \frac{d\mathbf{v}}{dt} = -\gamma \mathbf{v} + \mathbf{F} + \boldsymbol{\xi}(t)
\]

illustrates how the velocity (\( \mathbf{v} \)) of the Brownian particle evolves over time. Here, \( m \) represents the mass of the particle, \( \gamma \) denotes the friction coefficient, \( \mathbf{F} \) accounts for deterministic forces, and \( \boldsymbol{\xi}(t) \) represents the random force due to thermal fluctuations in the fluid. Additionally, the Einstein diffusion equation relates the diffusion coefficient (\( D \)) to the temperature (\( T \)) and the friction coefficient (\( \gamma \)), as follows:

\[
D = \frac{k_B T}{6\pi\gamma R}
\]

where \( k_B \) is the Boltzmann constant and \( R \) is the radius of the Brownian particle. This equation highlights the dependence of diffusion on temperature, friction, and particle size. Overall, Brownian motion plays a crucial role in various natural processes, including intracellular transport, diffusion within biological systems, and the dispersion of particles in fluids, contributing significantly to our understanding of stochastic processes and statistical mechanics.

\subsubsection{Simulation of Brownian Motion}

The location of a particle at time $t$ can be described using the following equations:

\[
\begin{aligned}
X_p(t) &= (X_p - \delta(t)) - \delta X(t) \\
Y_p(t) &= (Y_p - \delta(t)) - \delta Y(t) \\
Z_p(t) &= (Z_p - \delta(t)) - \delta Z(t)
\end{aligned}
\]

where $\delta(t)$ represents the step size. The speed of the Brownian motion is governed by random forces described by the random variables $\Delta X(t)$, $\Delta Y(t)$, and $\Delta Z(t)$, which follow a normal distribution with mean $0$ and variance $2D\Delta t$. Here, $D$ denotes the diffusion coefficient of the molecule, measured in units of $m^2/s^{-1}$, determining the rate at which the molecule moves.

\section{Measurement Campaign and Simulation}
\vspace{8pt}
The measurement campaign in this work is numerical: MATLAB is used to record the number of ABA molecules that reach a virtual receiver after their release from a root-side transmitter. The scenario is intended as an idealized molecular-communication representation of long-distance ABA movement through the xylem. Therefore, the results should be interpreted as a simulation-based study of diffusion-driven transport, not as a direct replacement for physiological measurements. Real xylem transport also includes sap flow, exchange with surrounding tissues, transporter activity, and ABA degradation; these effects can be added in later versions of the model.

\subsection{Simulation Scenario and Geometry}

The simulated plant geometry is shown in \FGR{fig:aba_transport_geometry}. The root region is treated as the transmitter (Tx), where ABA molecules are released after biosynthesis. The receiving region (Rx) represents the target soybean tissue reached by ABA after upward movement along the vascular path. A xylem segment of length $10~\mathrm{cm}$ is modeled as a bounded cylindrical channel, and the receiver is placed at the upper end of this path. In the receiver-size analysis, the spherical receiver radius is varied in the centimeter range to study the effect of target size on the number of detected molecules.

The number of released molecules is denoted by $Q$. Four release quantities are investigated, $Q=10^4$, $10^5$, $10^6$, and $10^7$. Using several values of $Q$ is useful because it separates two effects. First, a larger $Q$ increases the received signal amplitude because more molecules are available to reach the receiver. Second, a larger $Q$ reduces the relative Monte Carlo noise of the simulation, so the spatial cloud and the received signal become smoother and closer to the expected theoretical diffusion behavior.

\begin{figure}[ht]
    \centering
    \includegraphics[width=0.9\linewidth, height=0.75\linewidth]{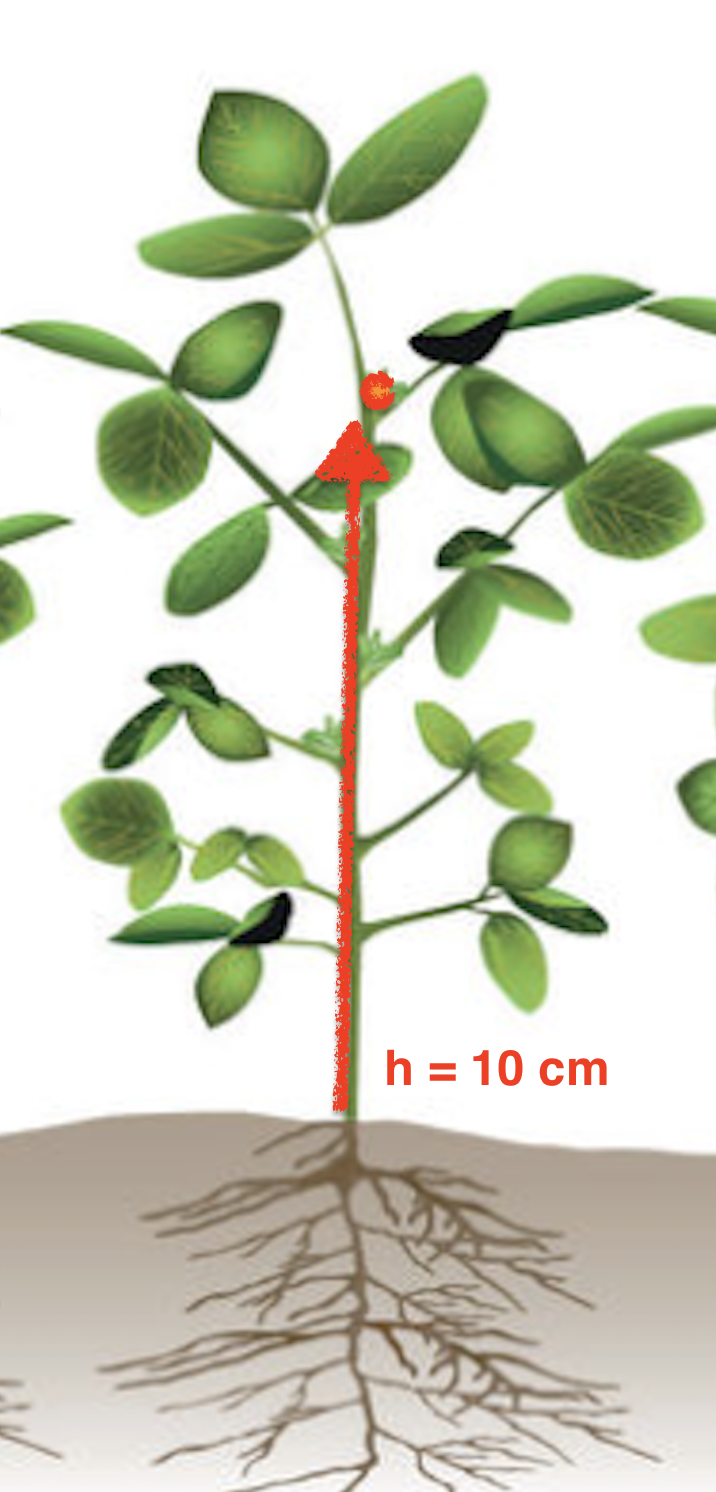}
    \caption{Geometry used for the ABA-transport simulation. The root side is modeled as the transmitter, while the target soybean tissue is represented as a spherical receiver located along a $10~\mathrm{cm}$ xylem path.}
    \label{fig:aba_transport_geometry}
\end{figure}

\subsection{MATLAB Diffusion Model}
\vspace{8pt}
The MATLAB implementation follows a particle-based Brownian-motion model. At each time step, every ABA molecule is displaced in the three Cartesian directions by a random increment with zero mean and variance controlled by the diffusion coefficient $D$ and the time step $\Delta t$:

\[
\Delta x,\Delta y,\Delta z \sim \mathcal{N}(0,2D\Delta t).
\]

Thus, for molecule $i$, the position update can be written as

\[
\mathbf{x}_i(t+\Delta t)=\mathbf{x}_i(t)+[\Delta x,\Delta y,\Delta z].
\]

The molecules are constrained to remain inside the cylindrical xylem domain. A molecule is counted as received when its distance from the receiver center is smaller than or equal to the receiver radius $r$:

\[
\left\|\mathbf{x}_i(t)-\mathbf{x}_{Rx}\right\| \leq r.
\]

The red point in the three-dimensional MATLAB plots indicates the transmitter position, the green point indicates the receiver center, and the blue points represent diffusing ABA molecules. The circular boundary visible in the plots corresponds to the xylem-like cylindrical domain used to confine the diffusion process.

\subsection{Three-Dimensional Diffusion Snapshots}
\vspace{4pt}
\FGR{fig:xyz_q4}, \FGR{fig:xyz_q5}, and \FGR{fig:xyz_q6} show spatial snapshots for $Q=10^4$, $10^5$, and $10^6$, respectively. For $Q=10^4$, the molecular cloud is sparse; individual random-walk events are clearly visible, and the received signal is expected to be highly fluctuating. When $Q$ is increased to $10^5$, the cloud becomes denser and the cylindrical shape of the bounded transport domain is easier to observe. At $Q=10^6$, the simulation produces a much smoother concentration pattern, which is consistent with the law of large numbers: when more particles are simulated, the random-walk ensemble better approximates the continuous diffusion equation.

These figures also clarify that increasing $Q$ does not change the diffusion physics or make a single molecule move faster. Instead, it increases the number of molecules available for detection and improves the statistical reliability of the simulated molecular distribution. This behavior is important for interpreting the received-molecule curves in following plots.

\begin{figure}[!ht]
    \centering
    \includegraphics[width=0.9\linewidth, height=0.68\linewidth]{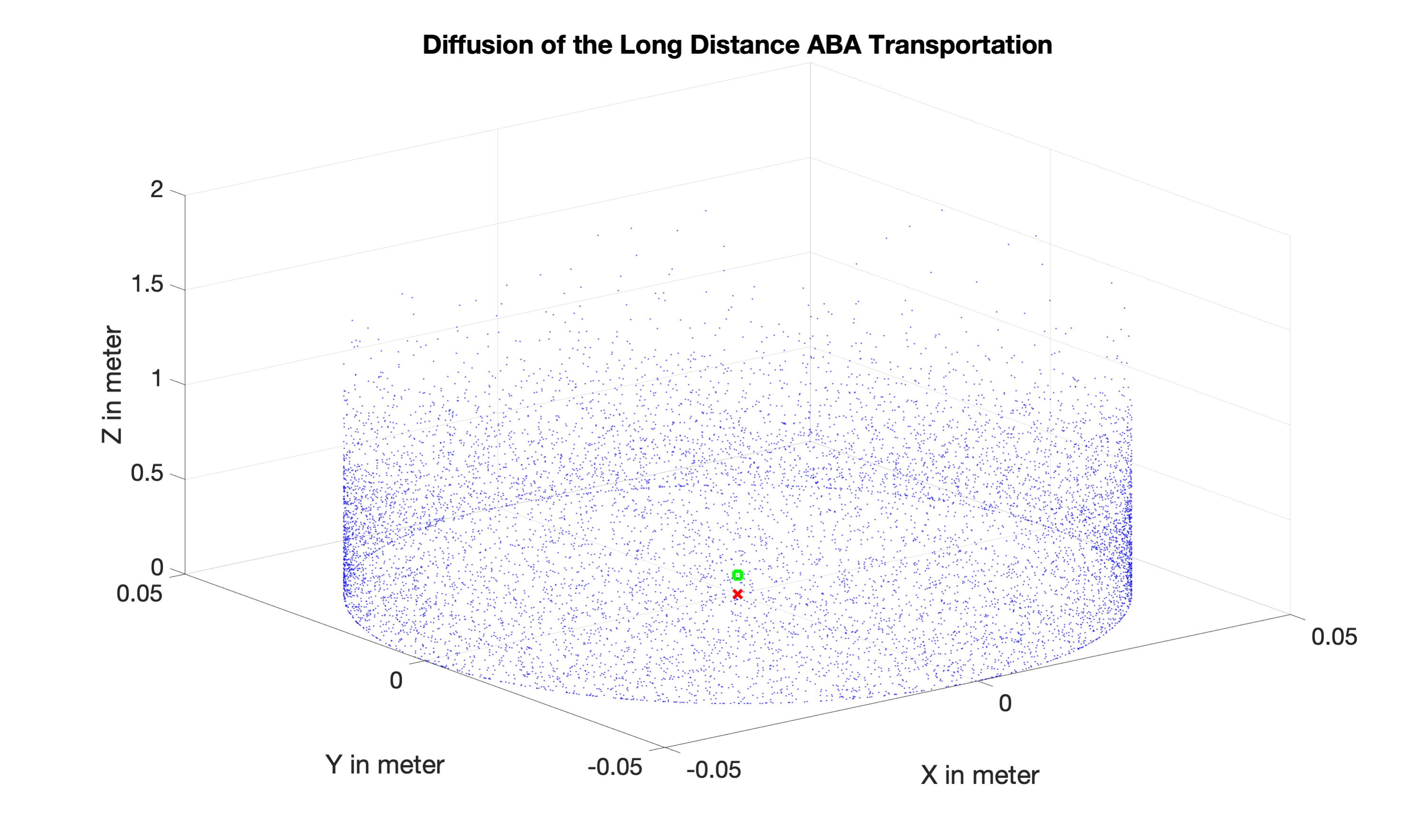}
    \caption{Three-dimensional diffusion snapshot for $Q=10^4$. The sparse molecule distribution leads to strong statistical fluctuations at the receiver.}
    \label{fig:xyz_q4}
\end{figure}
\begin{figure}[!ht]
    \centering
    \includegraphics[width=0.9\linewidth, height=0.68\linewidth]{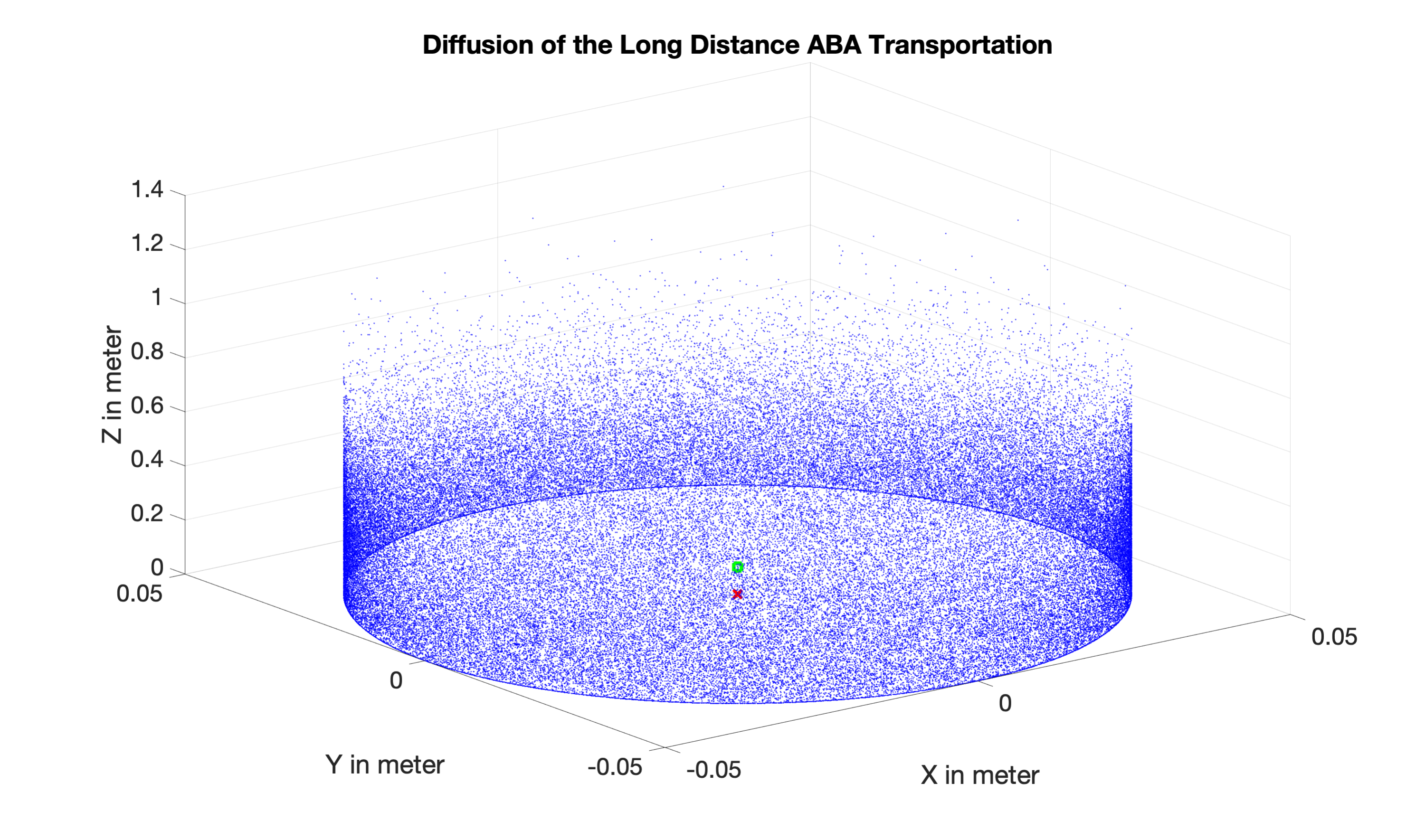}
    \caption{Three-dimensional diffusion snapshot for $Q=10^5$. The denser cloud provides a clearer representation of the bounded xylem-like transport domain.}
    \label{fig:xyz_q5}
\end{figure}
\begin{figure}[!ht]
    \centering
    \includegraphics[width=0.9\linewidth, height=0.68\linewidth]{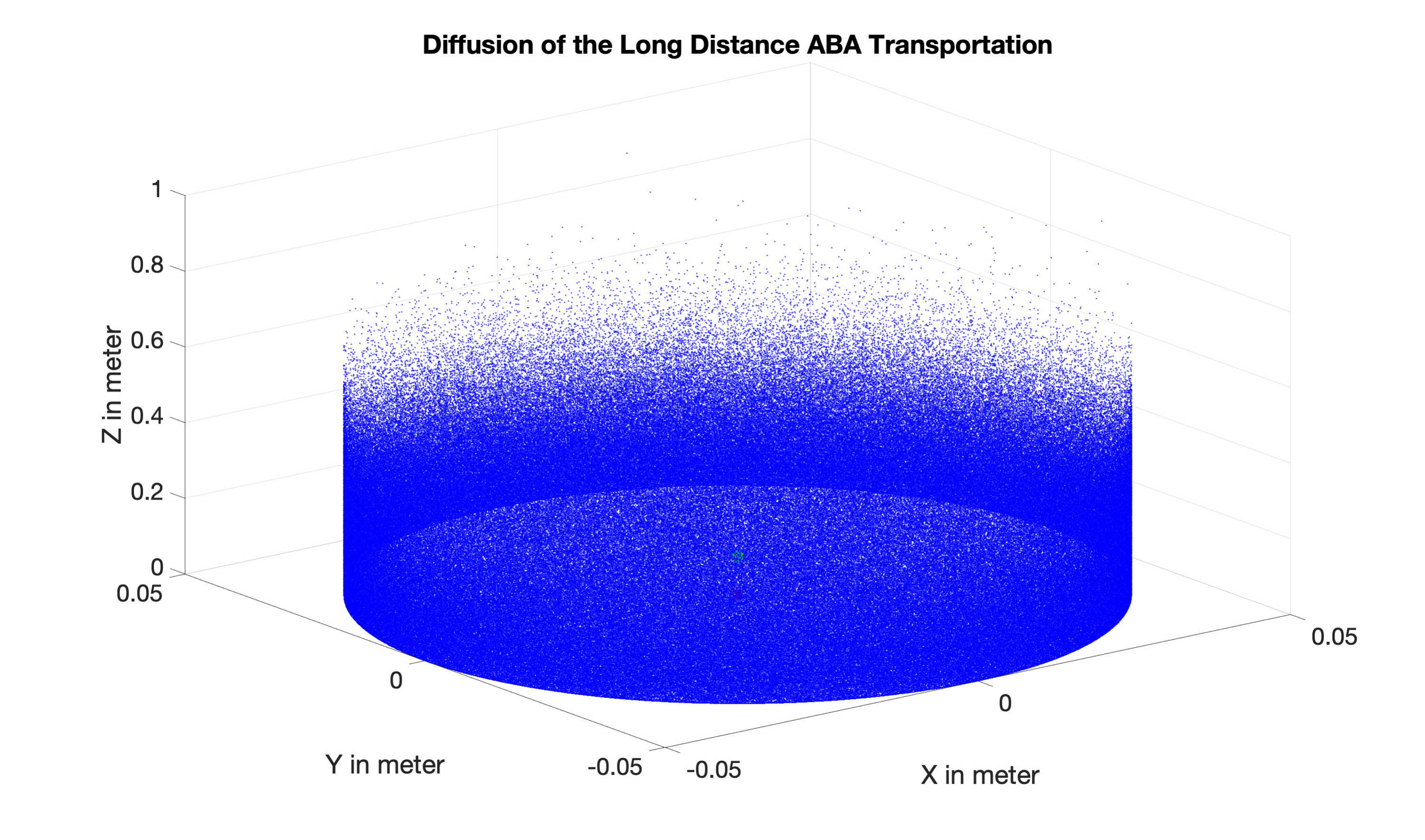}
    \caption{Three-dimensional diffusion snapshot for $Q=10^6$. The distribution becomes smoother and more consistent with the expected diffusion profile.}
    \label{fig:xyz_q6}
\end{figure}

\subsection{Received-Molecule Curves for Different Release Quantities}
\vspace{8pt}

\FGR{fig:capture_q4}--\FGR{fig:capture_q7} report the MATLAB receiver output for $Q=10^4$ to $Q=10^7$. The horizontal axis is time in seconds; in the MATLAB plots, the axis multiplier $10^8$ indicates that the displayed interval from 0 to 2 corresponds to a simulated interval from 0 to $2\times10^8~\mathrm{s}$. This very large time scale is a consequence of using a diffusion-only abstraction for long-distance transport. In real plants, advective xylem flow would usually accelerate hormone transport compared with pure Brownian diffusion.

The vertical axis should be interpreted as the number of molecules detected at each sampling instant, or within each sampling window, rather than the cumulative number of molecules ever captured. This interpretation is necessary because the plotted traces can decrease with time. A cumulative arrival curve would be non-decreasing. Therefore, the peaks in these plots represent moments when the receiver volume contains, or captures during a sampling interval, a large number of ABA molecules; the subsequent decrease indicates that the molecular cloud has spread away from the receiver region or has become more diluted over full domain.

\begin{figure}[ht]
    \centering
    \includegraphics[width=0.99\linewidth, height=0.75\linewidth]{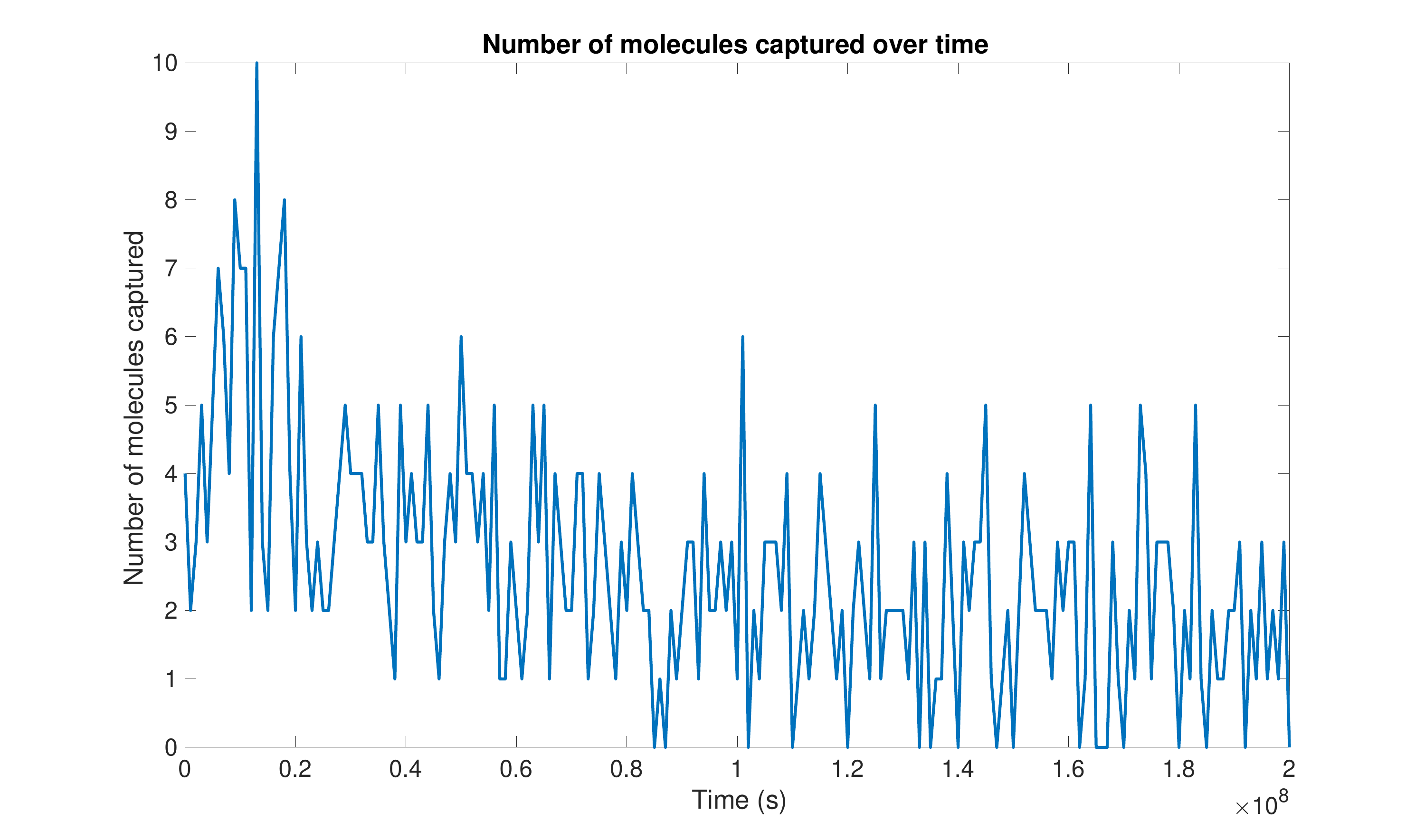}
    \caption{Receiver output $Q=10^4$. Low molecule count produces a noise.}
    \label{fig:capture_q4}
\end{figure}

For $Q=10^4$ in \FGR{fig:capture_q4}, only a small number of molecules reach the receiver, so the curve is dominated by random fluctuations. For $Q=10^5$ in \FGR{fig:capture_q5}, the received signal increases to several tens of molecules and shows a clearer decay trend after the initial high-arrival interval. 

\begin{figure}[ht]
    \centering
    \includegraphics[width=0.99\linewidth, height=0.75\linewidth]{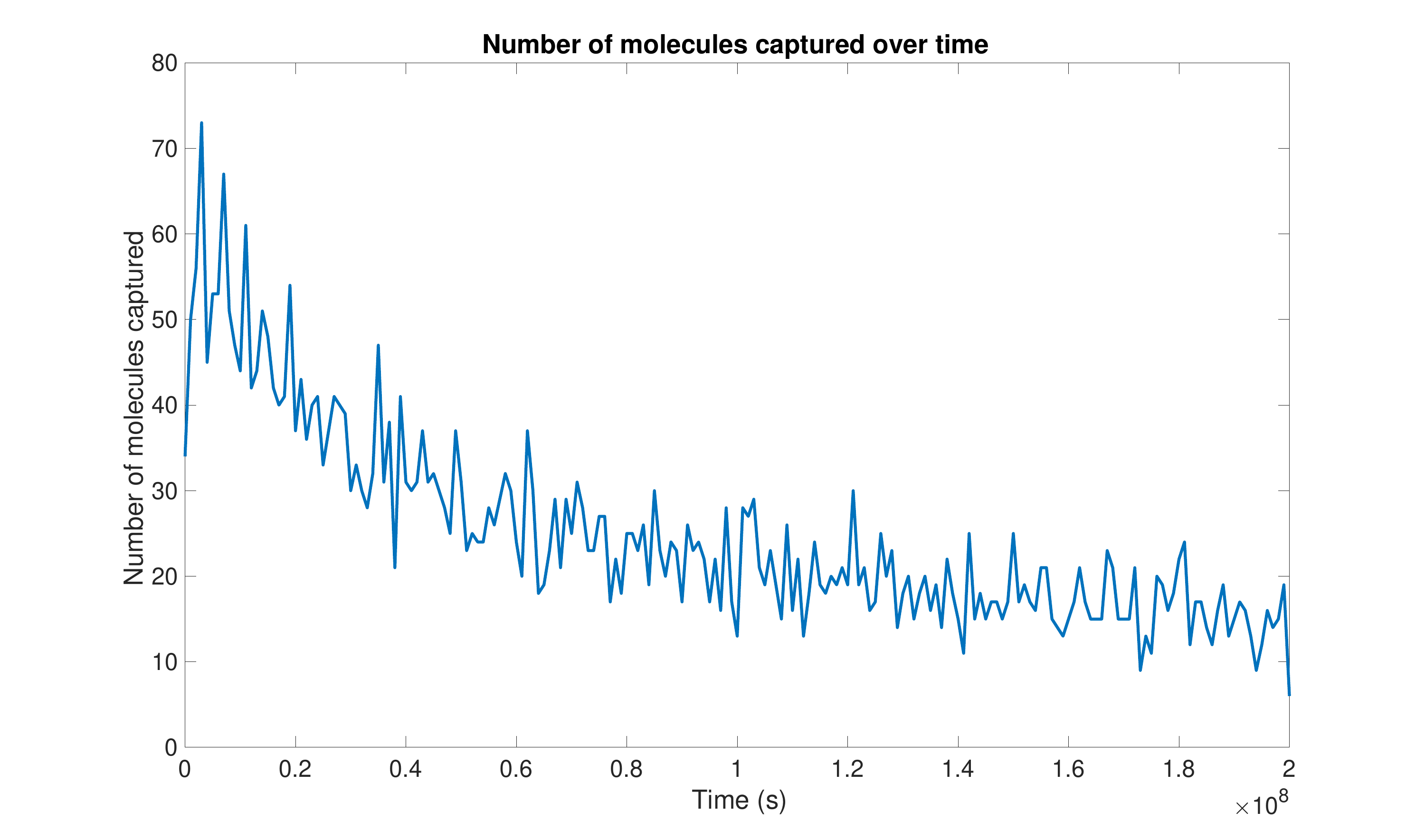}
    \caption{Receiver output for $Q=10^5$. The signal amplitude increases, while random fluctuations remain visible.}
    \label{fig:capture_q5}
\end{figure}

For $Q=10^6$ in \FGR{fig:capture_q6}, hundreds of molecules are detected, and the curve is smoother because the random fluctuations are averaged over a larger particle population. Finally, for $Q=10^7$ in \FGR{fig:capture_q7}, the received signal reaches the order of thousands of molecules, with an initial peak of approximately $5.5\times10^3$ molecules followed by a gradual decay. It demonstrates the expected scaling of the receiver response with the release quantity $Q$.

\begin{figure}[ht]
    \centering
    \includegraphics[width=0.99\linewidth, height=0.75\linewidth]{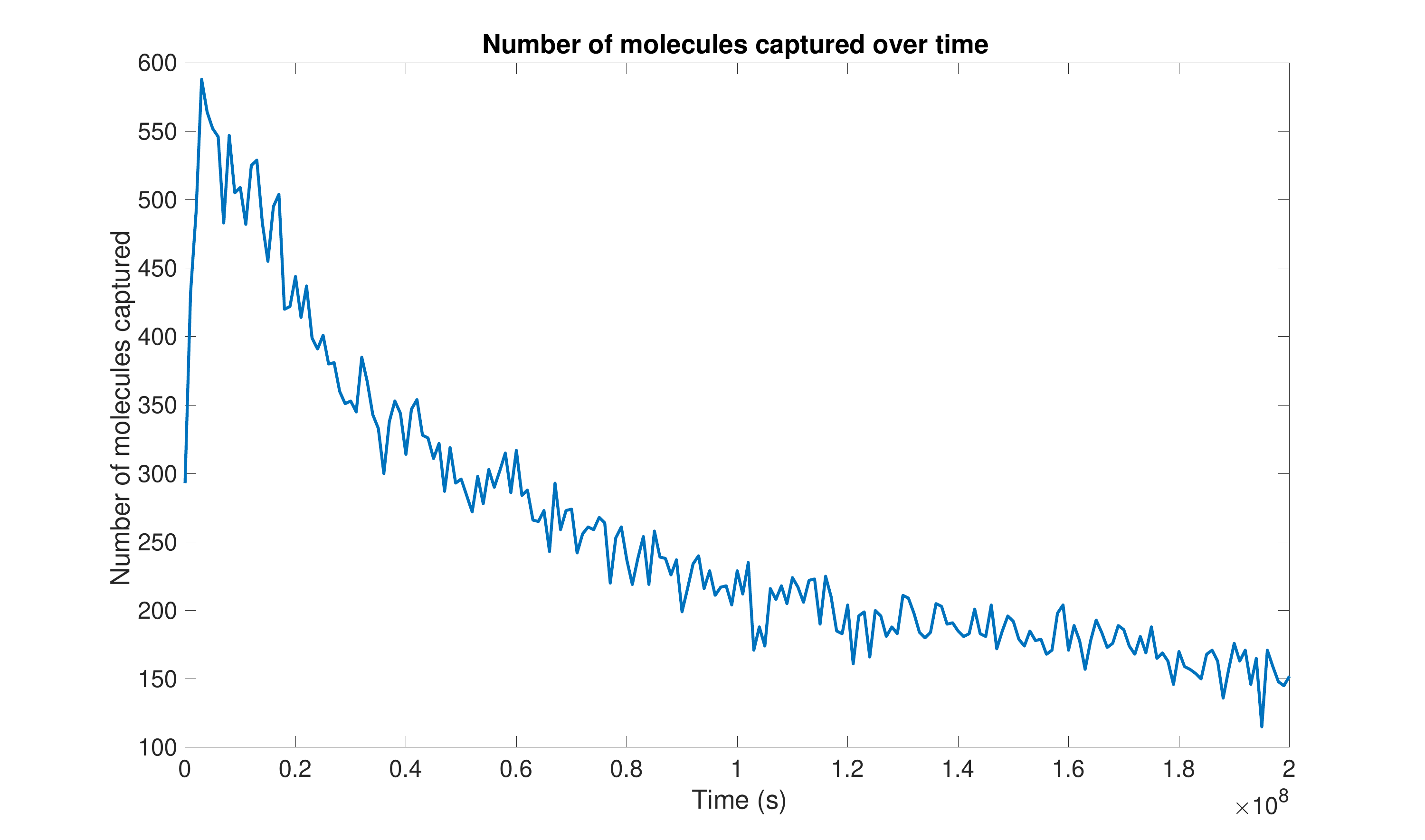}
    \caption{Receiver output for $Q=10^6$. A larger release quantity produces a smoother and stronger received signal.}
    \label{fig:capture_q6}
\end{figure}

The pulse-like behavior in \FGR{fig:capture_q7} is also consistent with the MATLAB release settings. The effective release is delayed and has a finite duration, so the receiver observes a strong early response when the densest part of the molecular cloud reaches the receiver. After the release interval, diffusion spreads the molecules over a larger volume, reducing the number detected inside the receiver region.

\begin{figure}[!ht]
    \centering
    \includegraphics[width=0.99\linewidth, height=0.8\linewidth]{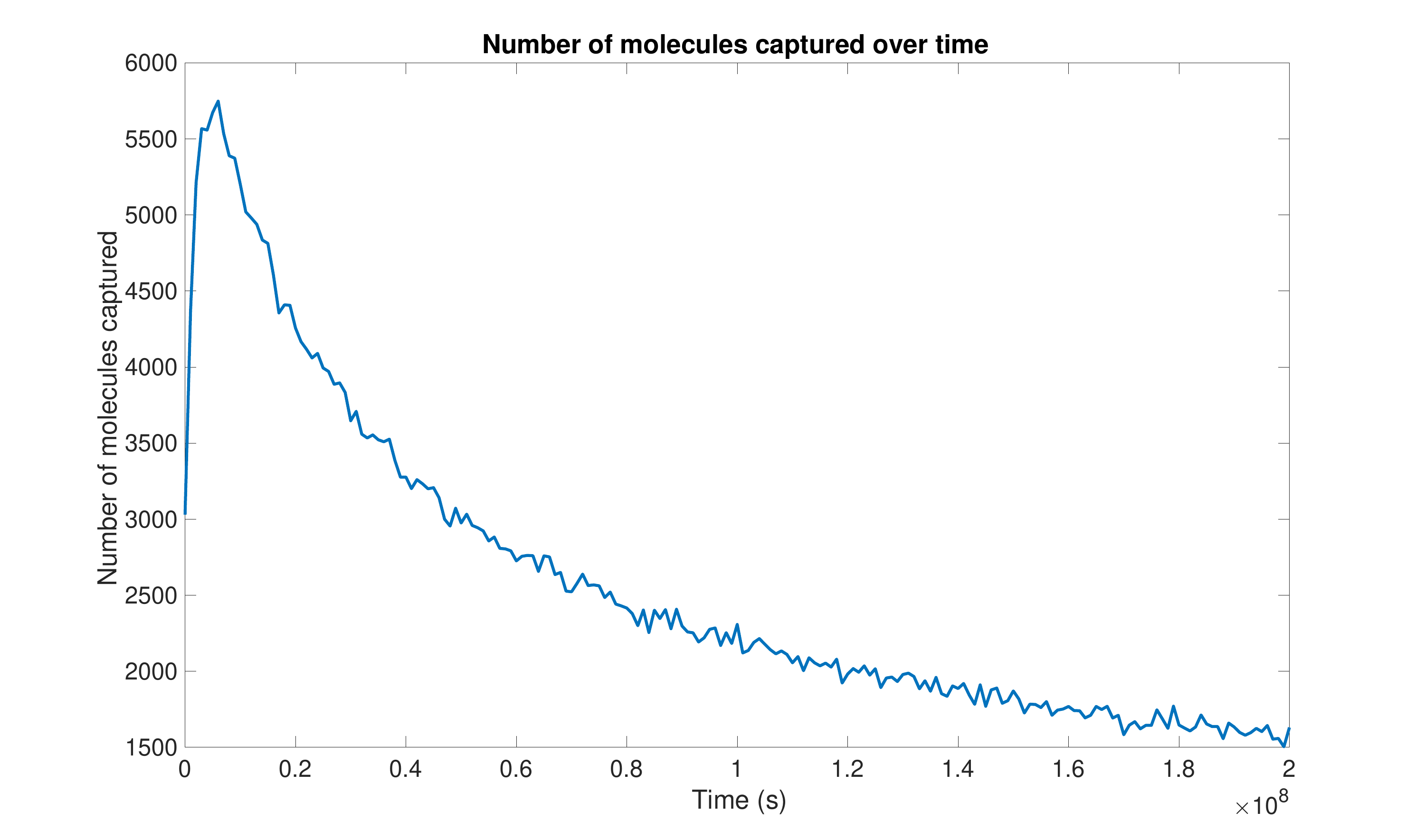}
    \caption{Receiver output for $Q=10^7$. The response has a high initial peak followed by a gradual decay as molecules spread through the domain.}
    \label{fig:capture_q7}
\end{figure}

\subsection{Effect of Receiver Radius}
\vspace{8pt}
\FGR{fig:receiver_radius} compares the received signal for different receiver radii while keeping the release quantity fixed at $Q=10^6$. The MATLAB legend uses $r=0.01$, $0.02$, and $0.03~\mathrm{m}$, corresponding to receivers with radii of 1, 2, and $3~\mathrm{cm}$. As expected, the larger receiver captures more molecules because it occupies a larger volume and presents a larger target region to the diffusing molecular cloud. The difference between the curves is therefore a direct geometric effect: increasing $r$ increases the probability that a molecule falls inside the receiver region at a given sampling instant.

This result is useful for the biological interpretation of the model. A larger target tissue, or a tissue with more effective uptake sites, should produce a stronger abscisic acid detection signal than a smaller target region under the same release conditions. In the communication analogy, the receiver radius plays a role similar to receiver aperture or capture cross-section: larger receivers improve the probability of detection and reduce the relative impact of stochastic fluctuations.

\begin{figure}[ht]
    \centering
    \includegraphics[width=0.99\linewidth, height=0.82\linewidth]{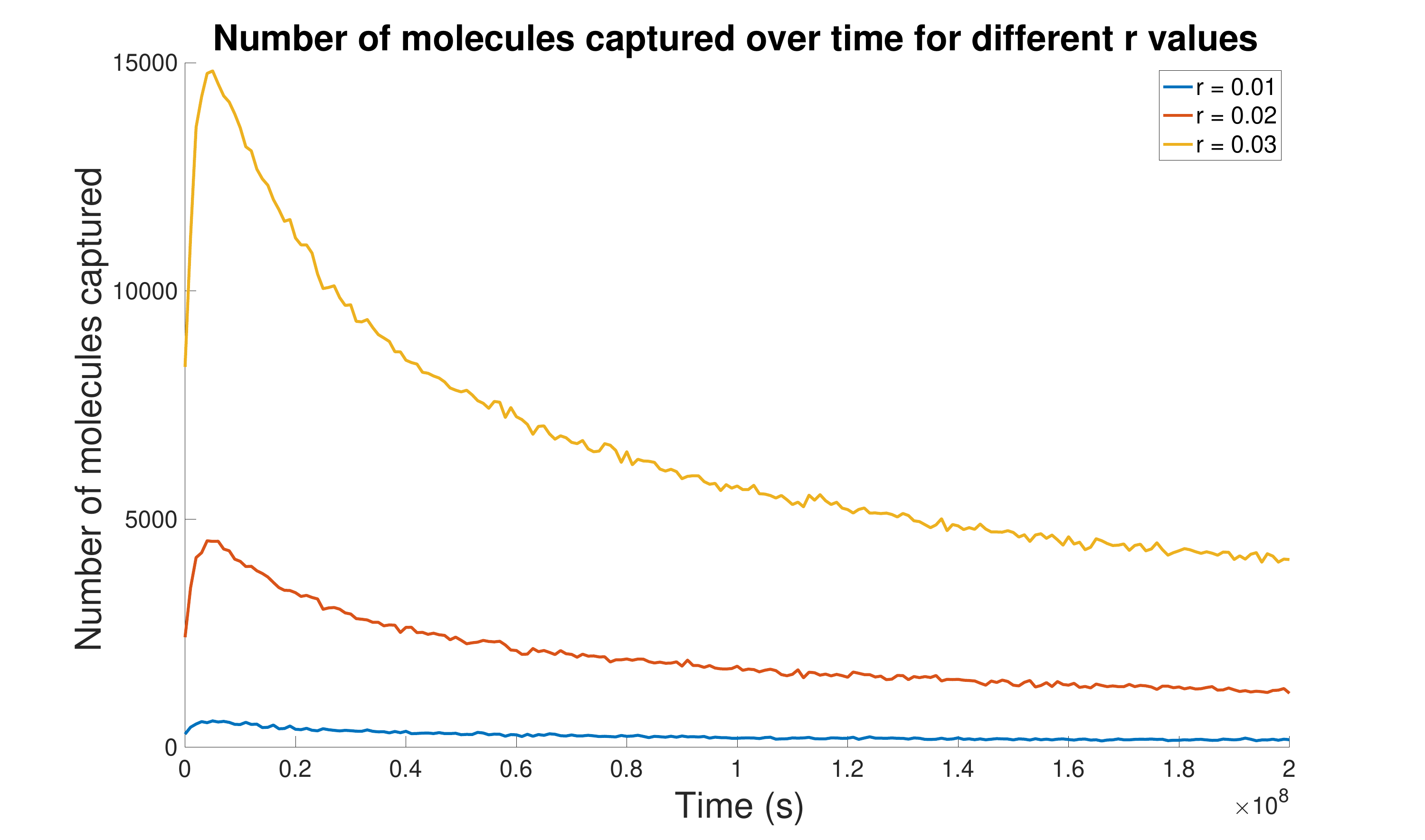}
    \caption{Effect of receiver radius for $Q=10^6$. Larger receiver radii capture more molecules and produce stronger received signals.}
    \label{fig:receiver_radius}
\end{figure}

\subsection{Summary of Simulation Findings}
\vspace{4pt}
The MATLAB plots support three main observations. First, the spatial diffusion snapshots become smoother as $Q$ increases, confirming that higher molecule populations reduce Monte Carlo noise. Second, the receiver-output curves scale with the number of released molecules: increasing $Q$ from $10^4$ to $10^7$ increases the detected signal from a noisy low-count response to a smooth high-amplitude response. Third, the receiver-radius comparison shows that geometry strongly affects detection, because larger spherical receivers intercept a greater fraction of the diffusing abscisic acid molecules. These observations validate the simulation as a useful first-order tool for studying abscisic acid transport as a molecular communication process, while also highlighting the need for future extensions that include xylem advection, biochemical degradation, and tissue-specific uptake mechanisms.

\section{Conclusion}
\vspace{4pt}
In summary, our investigation into the transport of abscisic acid (ABA) in plants has provided valuable insights into the complex mechanisms governing long-distance signaling in response to environmental stress. The first section focused on stimuli that elevate ABA production, with drought emerging as a prominent trigger. However, further exploration into additional stressors and their interplay could deepen our understanding of ABA-mediated stress responses, offering potential applications in crop improvement \cite{b15,b16,b17,b18,b19}. 

Moving to the production site of ABA, we identified specific tissues and cells as key players in ABA biosynthesis. This not only enhances our knowledge of ABA dynamics but also presents opportunities for targeted genetic interventions. By unraveling the regulatory networks governing ABA production in response to stress, we open avenues for precision breeding to enhance stress tolerance in crops. The decision on whether to employ xylem or phloem for ABA transport added a layer of complexity to our investigation. This choice seems to be context-dependent, influenced by the nature of the stressor. Exploring the factors dictating this preference could provide crucial insights into the adaptive strategies plants employ. Additionally, investigating crosstalk between ABA and other signaling molecules within the vasculature may uncover integrated stress response networks.

The experimental quantification of ABA levels grounded our theoretical framework in empirical evidence. The diversity of experimental approaches used for quantification not only validated our findings but also highlighted the need for standardized protocols across research groups. Advancements in imaging and omics technologies will further refine our ability to capture the dynamic nature of ABA responses in plants. The modeling of ABA transport in plants served as a crucial synthesis of our findings. These computational models offer a predictive framework for future experiments and agricultural applications. As models evolve, incorporating data from various plant species and environmental conditions will enhance their accuracy. Collaborations between experimentalists and modelers will be essential for refining and validating these models against real-world observations. Looking ahead, the implications of our research extend beyond scientific understanding. The knowledge gained from dissecting ABA transport mechanisms holds promise for addressing global challenges, such as climate change and food security. Applications in agriculture, inspired by our findings, could lead to the development of resilient crops capable of withstanding diverse environmental stresses. In conclusion, this interdisciplinary exploration of ABA transport in plants not only deepens our understanding of plant physiology but also lays the groundwork for sustainable and resilient agricultural practices in a changing world.

The measurement and simulation campaign further emphasized the importance of connecting biological assumptions with quantitative transport behavior. By representing ABA release from the root as a molecular transmitter, the xylem pathway as a bounded cylindrical channel, and soybean tissues as spherical receivers, the MATLAB model provided an intuitive bridge between plant physiology and molecular communication theory. The resulting plots showed that the initial number of emitted molecules, the diffusion time, and the receiver radius strongly affect how many ABA molecules are captured at the target site. Higher molecule quantities produced smoother and more reliable reception trends, while larger receiver radii increased the probability of capture. These observations indicate that future simulation work should incorporate additional biological factors, such as xylem sap flow, ABA degradation, active membrane transporters, tissue-specific storage, and experimentally measured diffusion coefficients. Including these parameters would make the model more realistic and would allow it to support predictions of ABA signaling strength under drought, salinity, cold, and other stress conditions.

\balance

\end{document}